\documentclass[12pt]{article}
\usepackage{graphicx}
\usepackage{aasms4}

\def\lapprox{\hbox{\lower .8ex\hbox{$\,\buildrel < \over\sim\,$}}}
\def\gapprox{\hbox{\lower .8ex\hbox{$\,\buildrel > \over\sim\,$}}}

\begin{document}

\title{Surviving companions of Type Ia supernovae: theory and observations}

\author{Pilar Ruiz--Lapuente$^{1,2}$}

\altaffiltext{1}{Instituto de F\'{\i}sica Fundamental, Consejo Superior de 
Investigaciones Cient\'{\i}ficas, c/. Serrano 121, E--28006, Madrid, Spain}
\altaffiltext{2}{Institut de Ci\`encies del Cosmos (UB--IEEC),  c/. Mart\'{\i}
i Franques 1, E--08028 Barcelona, Spain}

\begin{abstract}

\noindent
We review the theoretical background and the observational searches made for
surviving companions of Type Ia supernovae (SNe Ia). Theory comprises
 the characteristics
of the stellar binary companions of the exploding white dwarfs at the time of
the supernova outburst and the expected effects on them of the explosion, as 
well as their subsequent evolution. That includes space velocities, rotation, 
luminosities (with discussion of possible mechanisms producing very faint 
companions) . 

\noindent
We then present the searches already made in the Galactic remnants of Type Ia 
supernovae and we assess the results obtained up to now using ground--based
telescopes and the {\it Hubble Space Telescope} ({\it HST}). The same is done 
for 
the remnants of this type in the Large Magellanic Cloud.  We point 
to new SNRs of Type Ia that can be studied with groundbased telescopes,
the {\it HST} and the {\it James Webb Space Telescope} ({\it JWST}),
 using various approaches
such as characterization of peculiar stars through color--magnitude diagrams,
determination of their stellar parameters by spectral fitting, and astrometric
measurements. {\it Gaia} can provide, as well, useful astrometric information.
Most of these approaches 
 have been used in the SNe Ia remnants  already explored. 
The future goal is to enlarge the sample to determine which stellar systems 
do actually produce these explosions.

\end{abstract}

\keywords{Supernovae, general; supernovae, Type Ia; supernova remnants}

\section{Introduction}

\noindent
Type Ia supernovae (SNe Ia) are distance indicators that have 
made possible the discovery of the accelerated expansion of the Universe
(Riess et al. 1998; Perlmutter et al. 1999) and continue to be
 powerful probes for 
investigating the nature of dark energy.  SNe Ia arise as  thermonuclear
explosions of white dwarfs (WDs) in close binary systems, brought about by
mass gain of the WDs from their companion star.  
The nature of that companion star can either be a 
main sequence, subgiant, red giant,  
AGB, He star or another WD. The mode of mass gain can be through Roche--lobe 
overflow, stellar wind, merging or collision with a companion WD, merging
with the electron--degenerate core of a red giant. That bears also on 
the way explosive thermonuclear burning is ignited and how it propagates
inside the WD.

\smallskip
\noindent
A scenario (single degenerate scenario or SD) is that in which a WD 
grows close to the 
Chandrasekhar mass by accretion from a companion star and ignites
when the central density reaches $\rho$ $>$ 10$^{9}$ (Whelan \& Iben 1973; 
Nomoto 1982). The companion will survive in this case. 
The alternative is the double--degenerate scenario  (DD) which involves
 the merger of 
two electron degenerate objects, either two WDs (Webbink 1984; Iben \& Tutukov 
1984) or a WD and the core of an asymptotic giant branch (AGB) star 
(Livio \& Riess 2003; Soker 2013; 2014; 2019).
 In this case there is no surviving 
companion.  

\smallskip
\noindent
A number of reviews have dealt with the progenitors of Type Ia SNe
(Wang \& Han 2012; Livio 2013; 
 Maoz. Mannucci \& Nelemans 2014; Ruiz--Lapuente 2014)
and their 
explosion mechanisms (Hillebrandt \& Niemeyer 2000;
 Hillebrandt et al. 2013). 

\smallskip
\noindent
The bulk of the SNe Ia is best explained by the explosive ignition of a CO
 mixture, close to the center of a WD with a mass near to the Chandrasekhar 
mass. That is supported by many observations (Nomoto \& Leung 2017).
 The thermonuclear 
burning would propagate subsonically at first (deflagration) and make a 
transition to the supersonic regime (detonation), when reaching layers
at lower densities.

\smallskip
\noindent
Models involving CO WDs with masses below the Chandrasekhar mass have also
been proposed (see the review of Hillebrandt et al. 2013). The most viable 
mechanism, in this case, would be accretion of He on the surface of the WD, 
followed by its detonation. Compression of the CO core would, in turn, induce 
its detonation (Fink et al. 2010). It is not clear that this 
mechanism could fit the observed characteristics of most SNe Ia (Sim et al.
 2012).   

\smallskip
\noindent
Recently, observations of MUSSES1604D (Jiang et al. 2017) have lend some
support to the double detonation scenario, for a particular subtype of SNe Ia. 
The spectrum of an early red flash seems to be best explained by explosive 
thermonuclear burning close to the surface of the WD. A He WD might have been 
the mass donor and there is even the possibility that it survived the SN 
explosion (Shen \& Schwab 2017; Shen et al. 2018).

\smallskip
\noindent
Here, we 
especifically address the direct searches for surviving 
companions of SNe Ia in SN remnants and their theoretical background.  
In that, the information about the remnants is crucial for developing the
search of possible companions.

\smallskip
\noindent
Having said that, we know  about 300 SN remnants in our Galaxy (Green 2014), 
but their classification as
SN Ia remnant or as Core Collapse (CC) SN remnant is based on the 
charactheristics 
of their X--ray spectra. There are a couple of tens of SNRs classified 
unambiguously as SN Ia or CC SN.  This type of classification has been made
as well for a number of SNRs in the Large Magellanic Cloud. 
Four SNe Ia correspond to well known observed events in our Galaxy:
SN 1604 (known as Kepler's SN), SN 1572 (Tycho's SN), SN 1006, and SN 185 
(thought to correspond to the SNR RCW 86). The SN 1885A, in the Andromeda 
galaxy, has also been classified as SN Ia and its remnant located and studied.

\smallskip
\noindent
Detailed searches for companions in SNe Ia remnants were only conducted 
after 1997. Specifically, data were collected to 
  look for peculiar velocities, excess luminosities, 
spectroscopic and chemical anomalies in the stars located in the central
regions of recently produced, nearby SNRs of the Ia type. These
searches were performed in order to either 
identify such companions or discard their presence (early work by
 Ruiz--Lapuente 1997; 
Ruiz--Lapuente et al 2004). Tycho's SNR was first
to be explored and to be subsequently investigated in a number of 
papers (Ruiz--Lapuente 
et al. 2004; Ihara et al. 2007; Gonz\'alez Hern\'andez et al. 2009; 
Kerzendorf et al. 2009, 2013, 2018a; Bedin et al. 2014; Ruiz--Lapuente et 
al. 2019).
 The remnants of SN 
1006 (Gonz\'alez Hern\'andez et al. 2012; Kerzendorf et al. 2012, 2018b) and 
of Kepler's SN (Kerzendorf et al. 2014; Ruiz--Lapuente et al. 2018) have 
followed. In the LMC, five SNRs have been explored to some extent (Schaefer \& 
Pagnotta 2012; Edwards, Pagnotta \& Schaefer 2012; Pagnotta \& Schaefer 2015; 
Li et al. 2017, 2019).     

\smallskip
\noindent
This paper is organized as follows: In Section 2, 
 we first review our theoretical knowledge
of the physical features that should or might characterize the surviving
companions of SNe Ia (including some progenitor evolution scenarios that 
result in unusual survivors). In section 3, 
 we  describe the work that has been done up to now exploring the
Galactic SNRs attributed to SNe Ia. We also discuss
  what has been achieved on the 
SNRs in the LMC. The last Section summarizes 
what we have learned thus far about the SNe Ia companions.

\section{Theoretical predictions for surviving companions}

\smallskip

\noindent
There are three main areas for exploring the nature of surviving 
companions of SNe Ia: 1) evolution up to the explosion of the WD; 
2) interaction of the ejecta with the WD, and 3) the (long--term) 
post--impact evolution of the companion. 

\smallskip
\noindent
The first and the last one lie in the realm of (binary) stellar 
evolution codes while the second one is mostly explored using 
hydrodynamical simulations.

\smallskip
\noindent 
There have been several studies in the last decade that explore some
parts of the parameter space of the whole process from co--evolution
of the binary system to the final fate of the companion.

\subsection{Luminosities of possible surviving companion stars}

\smallskip

\noindent
There have been analytical studies of the impact of a 
supernova explosion on a binary companion (Wheeler, McKee \& Lecar 1974; 
Wheeler, Lecar \& McKee 1975), using a polytropic model for the companion 
star and a plane--parallel approximation for the structure of the 
supernova ejecta.

\noindent
The first detailed hydrodynamical (2D) simulations of 
the impact of SN Ia ejecta on  
a companion star, considering companions of different types (main sequence, 
subgiant, red giant), and estimating the post--impact luminosities, have 
been those of Marietta, Burrows \& Fryxell (2000). They have been followed by 
those of Pakmor et al. (2008) (3D; main sequence companion (MS) only), of Pan, 
Ricker \& Taam (2012a,b; 2014) (2D and 3D; main sequence, 
red giant (RG) and He star 
companions), and of Liu et al. (2013) (3D; main sequence companion).

\smallskip 
\noindent
The post--impact 
evolution of the companion star can only be followed at most up 
to the time when quasi--hydrostatic equilibrium is restored (on the order
of hours to days) due to the limitations of hydrodynamical simulations.
 The star is then 
still far from thermal equilibrium. However, potential companion candidates, 
found in nearby SNe Ia remnants, are several hundred years old 
(the youngest of which, Kepler SNR, is now more than 400 years old).

\noindent
In order to compare the hydrodynamical simulations of the impact on a
companion star with observations, a stellar 
evolution code has to be used to cover the time interval between explosion and
observations.
 Marietta, Burrows \& Fryxell (2000) mostly speculated about the 
long--term evolution of the companion.

\smallskip
\noindent
According to Marietta, Burrows \& Fryxell (2000), main sequence companions, 
while out of thermal equilibrium, might reach a luminosity of 500--5,000 
L$_{\odot}$, with a cooling timescale of 1,100--1,400 yr. After thermal 
equilibrium is restablished, they would return to the main sequence more
slowly, and continue the evolution corresponding to their mass (now reduced 
by the interaction with the SN ejecta). Subgiants would have a similar 
evolution, returning in a long term to a 
post--main--sequence track with a slightly lower 
luminosity than before impact.

\smallskip
\noindent
In the red giant case, the degenerate core is left surrounded by a hot, H/He
atmosphere. The residual envelope, with a small mass, when settling back down
to a thin layer, might reignite H shell burning. The star should evolve 
away from the red giant branch, along a track of constant luminosity and
increasing effective temperature, on a timescale of 10$^{5}$--10$^{6}$ yr. 
Marietta, Burrows \& Fryxell (2000) emphasize that the post--impact evolution 
should be handled with a 2D code, able to deal with initial models that are 
asymmetric in density and temperature and out of thermal equilibrium.    

\bigskip
\noindent
The long--term evolution of the companion star has been first calculated by 
Podsiadlowski (2003), for the subgiant case. It is a 1D calculation. 
To construct the initial model, mass is removed at a very high rate from 
a subgiant model in quasi--hydrostatic and thermal equilibria. A uniform 
heating source is then added to the remaining outer layers, and the 
subsequent re--equilibration of the star is followed. In this treatment of the
problem, the energy injected by the impact is deposited uniformly in the 
outermost 90\% radial extent of the star. The amount of that energy is a  
free parameter, and depending from its amount very different results are
obtained: a few centuries after the explosion the companion might appear 
from overluminous to underluminous, as compared with the pre--impact model.
An important result is that the re--equilibration of the outermost layers
 is much faster than 
the thermal timescale (i.e. the time to reach thermal equilibrium) 
 of the whole envelope of the initial subgiant.

\bigskip
\noindent
More recently, Shappee, Kochanek \& Stanek (2013) have explored the evolution 
of a main sequence companion of 1 M$_{\odot}$. They use the stellar evolution
 MESA (Modules for Experiments in Stellar Astrophysics) code (Paxton et al 2011)
 (1D) code.\footnote{http://mesa.sourceforge.net}
They first simulate the stripping of mass from
the star by an initial phase of rapid mass loss, but then they introduce 
a second phase when they add a heating source, which induces a wind that 
would simulate the mass loss by ablation (as in Podsiadlowski 2003). 
The internal energy added per unit
mass is made proportional to the ratio of the enclosed mass at a given radius
to the total mass of the model. Then they follow the subsequent evolution and
find that the star should remain overluminous (10 to 1,000 L$_{\odot}$) for 
1,000 to 10,000 yr.

\bigskip
\noindent
The most recent calculation of the post--impact evolution is that of Pan, 
Ricker \& Taam (2012b).
They use again the 1D MESA code, but starting from the results of their
own hydrodynamic simulations (Pan, Ricker \& Taam 2012a). The time steps 
are made very short initially, to closely follow the evolution when the 
star is still very far from equilibrium. They find that the evolution of 
the star not only depends on the amount of energy absorbed in the impact but 
also on the depth of the energy deposition.
They suggest that 
shock compression is an important factor (which has been ignored in 
some other studies) and they criticize the arbitrary setting of the depth
of energy deposition in previous work. 

\begin{table*}
\centering
\begin{minipage}{160mm}
\caption{The models in Pan, Ricker \& Taam (2012b). 
 (Courtesy of Kuo-Chuan Pan. @AAS. Reproduced with 
 permission.) }
\begin{tabular}{lrrrrrrrrrr}
\\
\hline
\hline
Model & $M_{i}$ & $P_{i}$ & $R_{i}$ & $L_{i}$ & $T_{eff,i}$ & $M_{f}$ & $P_{f}$ & $R_{f}$
 & $L_{f}$ & $T_{eff,f}$ \\
      & ($M_{\odot}$) & (day) &($R_{\odot}$) &($L_{\odot}$) & (K) & ($M_{\odot}$) & (day)
 & ($R_{\odot}$) & ($L_{odot}$) & (K) \\
\hline
A & 2.51 & 0.477 & 1.83 & 39.2 & 10696 & 1.88 & 0.350 & 1.25 & 2.35 & 6392 \\
B & 2.51 & 0.600 & 2.08 & 42.4 & 10224 & 1.92 & 0.466 & 1.50 & 3.64 & 6516 \\
C & 3.01 & 1.23 & 3.64 & 110.0 & 9800 & 1.82 & 1.09 & 2.63 & 8.06 & 6003 \\
D & 2.09 & 0.472 & 1.67 & 19.2 & 9358 & 1.63 & 0.353 & 1.19 & 2.09 & 6372 \\
E & 2.09 & 0.589 & 1.91 & 20.8 & 8933 & 1.59 & 0.470 & 1.42 & 3.15 & 6450 \\
F & 2.09 & 0.936 & 2.59 & 23.9 & 7934 &1.55 & 0.770 & 1.97 & 5.09 & 6182 \\
G & 2.00 & 1.00 & 1.70 & 17.6 & 9083 & 1.17 & 0.233 & 0.792 & 0.463 & 5355 \\
\hline
\end{tabular}

\medskip
The mass ($M_{\rm i}$), period (day), radius ($R_{\rm i}$), luminosity 
($L_{\rm i}$), and effective temperature ($T_{\rm eff,i}$) for different 
companion models at the beginning of RLOF for WD+MS systems, using the 
initial masses and orbital periods in Figure~7 of Hachisu, Kato \& Nomoto 
(2008), are given in columns 2--6. The mass ($M_{\rm f}$), period (day), radius 
($R_{\rm f}$), luminosity ($L_{\rm f}$), and effective temperature 
($T_{\rm eff,f}$) for companion progenitor models at the time of the SN 
explosion. See Figure 1.
\end{minipage}
\end{table*}

\begin{figure}
 \centering
 \includegraphics[width=0.8\columnwidth, angle=0]{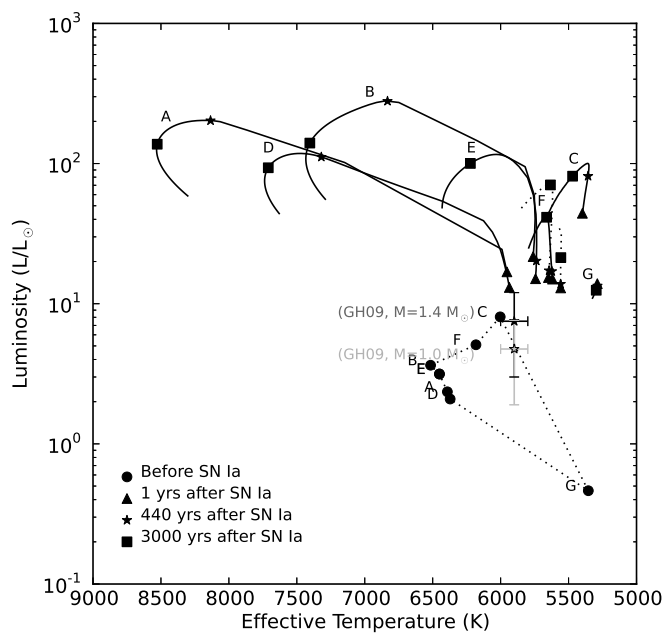}
 \caption{
Evolutionary tracks in the H--R diagram, for the different 
post--impact companion models in Table 1. Each track corresponds to the 
evolution for 10$^{4}$ yr of one of these models. Filled circles mark the
conditions of the stars just before the SN explosion. Filled triangles, those
$\sim$ 1 yr after explosion. Star symbols, 440 yr after it (the age of Tycho's
SN). Filled squares, 3000 yr after explosion. The star symbol with error
bars shows the observed luminosity and effective temperature of star G as 
measured by Gonz\'alez Hern\'andez et al. (2009). Pan, Ricker \& Taam (2012b).
 (Courtesy of Kuo-Chuan Pan. @AAS. Reproduced with 
 permission.) }
 \label{Figure 2} 
\end{figure}

\smallskip
\noindent
The post--impact companion stars rapidly expand on time scales $\sim$ 
10$^{2}$ -- 10$^{3}$ yr (depending on the pre--impact structure). Initially, 
the luminosity inside the envelope is much higher than the surface luminosity. 
Due to that, the outermost $\lapprox$ 1\% of mass expands and the luminosity 
profile in the outermost 10\% of mass flattens due to radiative diffusion.
The local thermal time scale in the envelope region is much shorter than the
global thermal time scale. After $\sim$ 200 yr, the deposited energy has been
radiated away and the star begins to contract, releasing gravitational energy.
The luminosity decreases and the effective temperature increases. It will
return to the Zero--Age Main Sequence ({\it ZAMS}) on a global thermal time
scale. The luminosities, 440 yr after the explosion (Figure 1), 
could be as low as 
$\approx$ 13 $L_{\odot}$, which is far below the predictions of Shappee, Kochanek
\& Stanek (2012). Interestingly, all models considered by Pan, Ricker \& Taam
(2012b) have, at the referred time, effective temperatures $T_{\rm eff} 
\gapprox$ 5300 K. In a later work, Pan, Ricker \& Taam (2014) present in 
more detail the observable characteristics of post--impacted  MS
 and He WD companions, 
such as their color and luminosity evolution. It is worth noticing that their
new class of companion models, the He WDs, as they were in a
 closer orbit with the exploding WD, gained higher velocities and reached
 effective temperatures on the range of $T_{\rm eff}$ from 10,000K to 70,000 K.

\subsection{Ejection velocities}

\noindent
Typically, the SNe Ia explosion will disrupt the binary system 
(in non--merging systems). 
The companion, its linear momentum unchanged, will start moving along a straight
line, with the orbital velocity it had at the time of the explosion. The SN 
ejecta, moving much faster, will overrun it and the portion intercepted by the 
companion will impart a kick, strip some mass from its outer layers, and
inject some energy into the material that remains bound (the latter inducing
extra mass ejection and puffing up). The escape velocity and kick might give 
the companion an unusual kinematic signature when compared with the 
surrounding stars. 
 Hydrodynamical simulations (Marietta, Burrows \& 
Fryxell 2000) of the interaction between ejecta and companion showed that
the kinematic effect of the kick (imparted perpendicularly to the orbital 
velocity) was minor as compared with the escape velocity. The momentum 
gained from the kick depended on orbital 
separation and on how compact the companion was. It ranged from 12\% to 
50\% of the momentum the star had before the explosion in the cases of 
main-sequence and subgiant companions, the kick being  much smaller in the 
case of red giants. Similar values have later been found by Pan, Ricker \& 
Taam (2014). 

\smallskip
\noindent
The orbital velocities depend on the nature of the companion star of the
WD. Since the companion is assumed to be filling its Roche lobe at the time 
of the explosion (save in the case of wind--accretion systems, where the 
WD accretes matter from the wind of a red--giant or supergiant star), 
the orbital separation $a$ should be smallest for main--sequence companions and
largest for red--giant ones. Using Eggleton's (1983) approximation for the 
radii of the Roche lobes:

\begin{equation}
R_{L} = a\left[{0.49\over 0.6 + q^{-2/3}{\rm ln}(1 + q^{1/3})}\right]
\end{equation}

\noindent
where $R_{L}$ is the Roche--lobe radius of the secondary star and $q \equiv 
M_{2}/M_{1}$; we see that for a fixed $q$, in order for the secondary to be
filling its Roche lobe the separation has to vary as the radius of the star. 
Then, assuming circular orbits, we have from Kepler's law that 

\begin{equation}
P^{2} = {a^{3}\over M_{1} + M_{2}}
\end{equation}

\noindent
where $P$ is the period in years of the binary, $a$ the orbital separation in
astronomical units, and $M_{1}$, $M_{2}$ are in solar masses. So we have that 
$v_{\rm orb} \propto a^{-1/2}$, and the main--sequence companions should move 
faster than the red--giant ones. 
If we look 
at the typical velocities  for main--sequence 
stars, subgiants and red giants obtained by Marietta, Burrows \& Fryxell (2000)
and Han (2008), those are of 250--300 km s$^{-1}$, 
150 km $s^{-1}$,  and  80 km s$^{-1}$ 
respectively. 

\smallskip
\noindent   
There is also the possibility, within the double--degenerate channel to
produce SNe Ia, that the explosion could be triggered just at the beginning
of the coalescence process of the two WDs, by detonation of a thin helium
layer coming from the surface of the less massive object, which would then
compress the more massive WD and detonate its core (Shen et al. 2018).The 
less massive WD would thus survive. In that case, the orbital velocity being 
very high ($>$ 1000 km s$^{-1}$), a hypervelocity WD would be ejected. Shen et 
al. (2018) claim that three objects found in the DR2 from the {\it Gaia} space 
mission are in fact such hypervelocity WDs. They do not look as typical WDs 
but they might result from heating and bloating of a WD companion.
WDs  with $M\gapprox 0.15$ M$_{\odot}$, however, do not become significantly 
bloated when heated. That can be seen from the WD cooling curves, since
the heating due to the impact of the supernova ejecta should follow the same
curves, now towards increasing temperatures. Only very low mass WDs are 
the outcome of stars resembling those hypervelocity objects.

\smallskip
\noindent
Since, in general, the peculiar velocity of a surviving companion will form
any angle with the line of sight to the SNR, it should be detected as both 
an excess in radial velocity and in proper motion, as compared with the average
of the stars at the same location within the Galaxy (or the host galaxy, in 
the case of extragalactic SNe Ia). A common problem when examining possible
candidates to SNe Ia companions is to determine their distances. Accurate 
distances to individual stars have to be obtained, most of the time, by 
deducing their spectral types and luminosity classes from their spectra and
then comparing the absolute magnitudes in different passbands with  
photometric measurements. The same spectra are used to measure the radial
velocities. Concerning proper motions, astrometric measurements from {\it 
Hubble Space Telescope} images taken at different epochs have been used.

\subsection{Rotation}

In close binary systems, tidal interaction tends to make rotation of the
component stars synchronous, that is, to make the rotation period equal to
the orbital period. The gravitational attraction of the companion star induces 
a tidal bulge on each component of the system (the same way as the Moon acts on 
the Earth's oceans). Synchronization is due to the viscosity of the fluid the 
star is made of. 

\smallskip
\noindent
When rotation is not synchronous, the tidal bulge creates a torque that either 
accelerates rotation when its period is longer than the orbital period or 
slows it down when it is shorter. 

\smallskip
\noindent
The timescale for synchronization, for a given initial system and any other 
influences on rotation (such as mass loss) being absent, thus only depends on 
the viscosity of the fluid: the higher the viscosity, the shorter the 
timescale. When the envelope of the star is in radiative equilibrium, only
radiative damping acts and the timescales are long. When the envelope is
convective, turbulent viscosity, much higher, is the relevant one and 
the timescales are much shorter.   

\smallskip
\noindent
Main sequence companions with masses $M \lapprox 1.3\ M_{\odot}$, subgiants, 
and red giants do posess surface convection zones. Therefore, we should 
expect that, in systems made of a white dwarf plus a companion of one of 
these types, from tidal interaction alone, and for times like those mediating 
between the epoch when the system becomes close and that of the SN Ia 
explosion, the rotation had become synchronous.

\smallskip
\noindent
We are ignoring here the loss of angular momentum due to mass loss (mass 
transferred to the white dwarf plus mass lost by the system) by the companion 
star, which should slow the rotation down. For fast mass 
losses, that would compete with the synchronization mechanism. 

\smallskip
\noindent
The rotation of the companion star post--explosion is likely not that
of pre--explosion but smaller, due to several effects. 
From all the existing hydrodynamic
simulations, mass is reduced by stripping (momentum imparted) and ablation
(energy deposition that unbinds layers that are below those removed by 
stripping). For a given total amount of mass lost, matter that was at some 
depth inside the star then becomes the new surface layer.  

\smallskip
\noindent
In all simulations, immediately after the explosion main sequence stars and 
subgiants do not just recover their former radii but are puffed up and do 
not return to the radii corresponding to their new masses and to their
evolutionary stages until thermal equilibrium is restored (between 
1400 and 11000 yr, in Marietta, Burrows \& Fryxell 2000). 
Red giants lose
almost all their envelope and the residual one also expands. When it 
recontracts, H shell burning, temporarily extinguished, could be reignited
and the star recover its former luminosity with a smaller radius (Marietta,
Burrows \& Fryxell 2000), although this is still mostly speculation by now. 
During this contraction phase the star would spin up.

\smallskip
\noindent
Liu et al. (2013) have made 3D Smoothed Particle Hydrodynamics (SPH) 
simulations of the impact of SN Ia ejecta on a main--sequence companion of 
1 M$_{\odot}$, using the GADGET--3 code (Springel et al. 2001; Springel 2005).
The evolution of the binary system previous to the explosion has been 
calculated with a 1D stellar evolution code (Eggleton 1973). Figure 2
displays the evolution in luminosity of the donor star and Figure 3 that of the
orbital velocity.

\begin{figure}
 \centering
 \includegraphics[width=0.8\columnwidth]{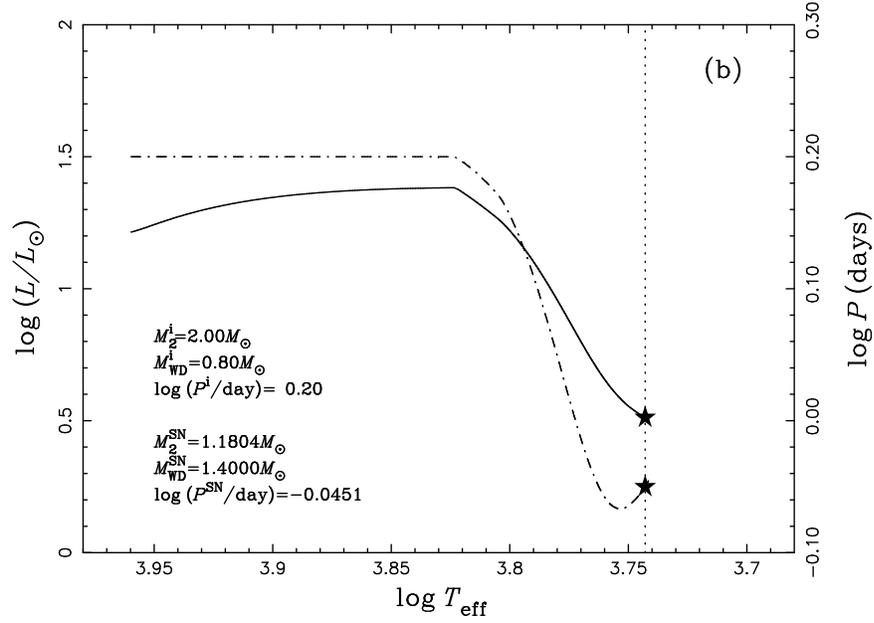}
 \caption{Evolution of the luminosity (left vertical axis) and the orbital 
 period (right vertical axis) of the donor star and of the system, 
 respectively, in one of the models calculated by 
 Liu et al. (2013), up to the point of explosion. The solid curve corresponds
to the evolution of the donor and the dash--dotted curve to that of the
period. (Courtesy of Zhen--Wei Liu. @Springer A \& A.
 Reproduced with 
 permission.)}
 \label{Figure 1}
\end{figure}

\smallskip
\noindent
In these simulations the rotational velocity of the companion star is 
significantly reduced to about 14\% to 32\% of its pre--explosion value
due to the expansion of the companion and because 35\% to 89\% of the 
initial angular momentum is carried away by the stripped matter (see Figure 3).
 Liu et al. (2013) also find that the radial distribution
of the rotation of the companion becomes approximately constant $\sim$10$^{4}$
s after the explosion (Figure 3).

\begin{table*}
\centering
\begin{minipage}{160mm}
\caption{The impact simulations of Liu et al. (2013).(Courtesy of 
Zhen--Wei Liu. @Springer A \& A.
 Reproduced with 
 permission.)}
\begin{tabular}{@{}lcccccccccc@{}}
\\
\hline
\hline
Model& $M_{2}^{SN}$ & $P^{SN}$ & $R_{2}^{SN}$ & $a^{SN}$ & $v_{rot}^{SN}$ & $v_{rot}^{f}$
 & $v_{rot}^{ff}$ & $J_{spin}^{SN}$ & $J_{spin}^{f}$ & $M_{bound}$ \\
     & [$M_{\odot}$] & [days] & [$R_{\odot}$] & [$R_{\odot}$] & \multicolumn{3}{c}{[km s$
^{-1}$]} & \multicolumn{2}{c}{[10$^{50}$ g cm$^{2}$ s$^{-1}$]} & [$M_{\odot}$] \\
\hline
MS--160 & 1.21 & 0.29 & 0.93 & 2.55 & 160 & 52 & 98 & 2.94 & 1.31 & 1.04 \\
MS--131 & 1.23 & 0.56 & 1.45 & 3.94 & 131 & 40 & 78 & 2.07 & 0.92 & 1.06 \\ 
MS--110 & 1.18 & 0.91 & 1.97 & 5.39 & 110 & 25 & 46 & 2.25 & 0.62 & 0.95 \\
MS--081 & 1.09 & 2.00 & 3.19 & 8.92 &  81 & 12 & 16 & 2.32 & 0.26 & 0.84 \\
\hline
\end{tabular}

\medskip
Here, $M_{\mathrm{2}}^{\mathrm{SN}}$, $P^{\mathrm{SN}}$,
$R_{\mathrm{2}}^{\mathrm{SN}}$, $a^{\mathrm{SN}}$, $v_{\mathrm{rot}}^{\mathrm{SN}}$ 
and $J_{\mathrm{spin}}^{\mathrm{SN}}$ are the mass, the orbital period, the radius,
the spin velocity and angular momentum of the companion star at the
moment of the explosion, respectively.
$v_{\mathrm{rot}}^{\mathrm{f}}$, $J_{\mathrm{spin}}^{\mathrm{f}}$ and
$M_{\mathrm{bound}}$ 
denote the spin velocity, the angular momentum, and
the total bound mass of the companion star after the SN impact. 
$v_{\mathrm{rot}}^{\mathrm{ff}}$ is the rotational velocity 
at the surface after the thermal equilibrium is restablished. Note
that the rotational velocity, 
$v_{\mathrm{rot}}^{\mathrm{SN}}$,
is calculated by assuming that the rotation of the star is locked 
with the orbital motion due to tidal interactions.
These four models are shown in Figure 3. 
\end{minipage}  
\end{table*}

\begin{figure}
 \centering
 \includegraphics[width=0.8\columnwidth, angle=0]{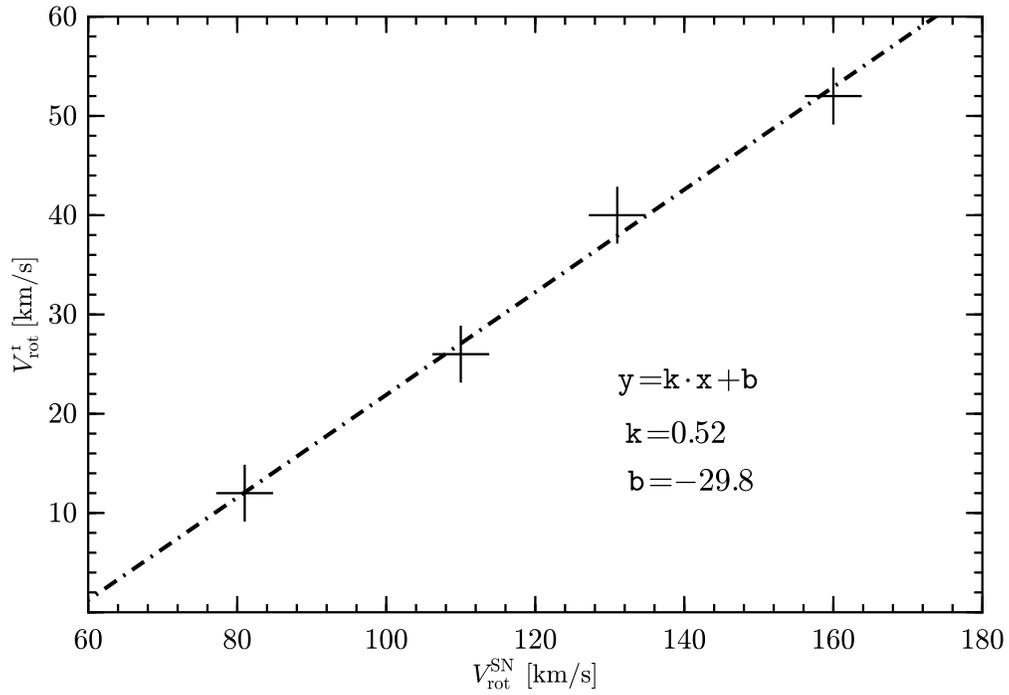}
 \caption{Initial rotational velocity at the time of the SN explosion $vs.$ 
rotational velocity after the explosion, for four different models from 
Liu et al. (2013). The crosses correspond to the four models of Table 2.
 (Courtesy of Zhen--Wei Liu. @Springer A \& A. Reproduced with 
 permission.)}
 \label{Figure 2} 
\end{figure}

\clearpage

\smallskip
\noindent
Pan, Ricker \& Taam (2012a,b; 2014) have made 3D hydrodynamic simulations of SN 
impacts
on different companion models, using the FLASH version 3 code (Fryxell et al. 
2000; Dubey et al. 2008). The progenitor systems were constructed and the 
post--impact evolution of the remnant star followed with MESA. 
The progenitor models were taken from 
Hachisu, Kato \& Nomoto (2008), who studied the binary evolution of WD+MS 
systems and found the region of the donor mass--orbital period plane where
SNe Ia can occur, but they were recalculated with the MESA code from the  
 ZAMS to the Roche--lobe overflow. Most companion stars of the
WD are slightly evolved MS stars at the time of the explosion, but some of
them were still close to the ZAMS. Their
 masses ranged between 2 and 3$M_{\odot}$. 

\smallskip
\noindent 
From the hydrodynamic simulations, the companion stars are heated and lose 
$\sim$10\%--20\% of their mass due to stripping and ablation by the SN ejecta. 
To follow the subsequent evolution, the resulting 3D models are turned into
1D models.The specific angular momentum is that obtained from the 3D 
simulations.

\begin{figure}
\centering
\includegraphics[width=0.8\columnwidth, angle=0]{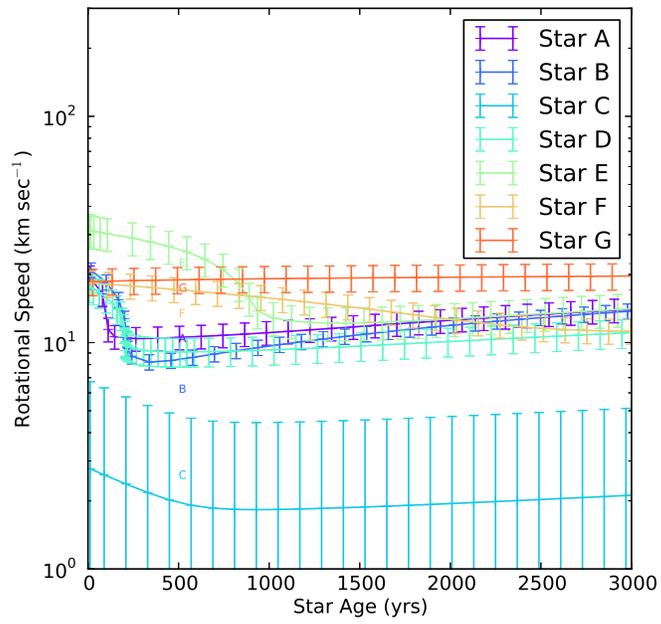}
\caption{Evolution of surface rotational speed for different companion 
          models considered by Pan, Ricker \& Taam (2012b) (see Table 1). 
(Courtesy of Kuo-Chuan Pan. @AAS. Reproduced with 
 permission.).}
\label{Figure 4} 
\end{figure}

\smallskip
\noindent
Figure 4 shows the evolution of the surface rotational speed for all the
remnant star models considered by Pan, Ricker \& Taam (2012b). We see that, 
after the impact, the rotation speed decreases as the post--impact remnant
star is expanding. In the most rapidly evolving models (stars A, B, and D),
the rotation speed falls to less than 10 km s$^{-1}$ within the first 
500 yr. Past 1000--1500 yr, the stars start to contract slowly, increasing
the surface rotation speed. Therefore, according to these calculations, the 
post--impact remnant stars do not need to be fast rotators even in the WD plus 
main--sequence systems.

\subsection{The spin--up/spin--down mechanism}

\smallskip
\noindent
In the ``classical'' SD channel (Whelan \& Iben 1973; see Introduction),
 accretion 
makes the mass of the WD grow until reaching the Chandrasekhar mass, 
$M_{\rm Ch}$, at which point the star begins to contract fast and explosive C 
burning is ignited at its center (or close to it). In that case, to the mass 
growth also corresponds gain of angular momentum by the WD. That would be the 
spin--up stage in the evolution of the binary system.  

\smallskip
\noindent
In a rotating WD, the critical mass, $M_{\rm crit}$, can considerably exceed 
$M_{\rm Ch}$. Ostriker \& Bodenheimer (1968) found, for an extreme case of 
differential rotation, that a mass as high as 4$M_{\odot}$  would still be 
stable, while Yoon \& Langer (2004, 2005) calculated that WDs could reach 
$\approx$ 2$M_{\odot}$ before exploding.

\smallskip
\noindent
Assuming that mass transfer stops at some point, due to exhaustion or 
contraction of the
companion's envelope, before the absolute stability
limit for the WD is reached, one would be left with a detached binary made of a 
rotating, super--Chandrasekhar mass WD plus an evolving companion. Then, 
rotation might start to slow down, until the point is reached where the 
decreasing $M_{\rm crit}$ becomes equal to the actual mass of the WD. That 
would be the spin--down stage.

\begin{figure}
\centering
\includegraphics[width=0.8\columnwidth,angle=0]{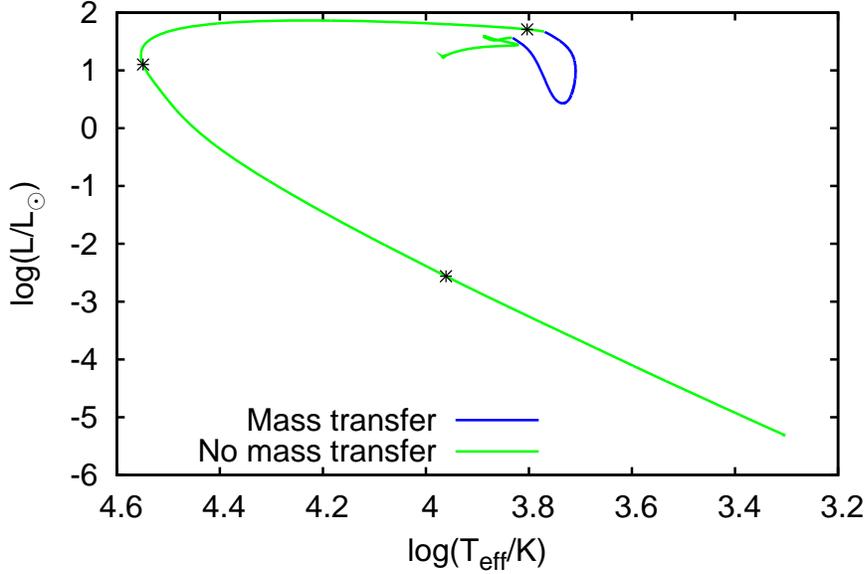}
\caption{
Hertzsprung--Russell diagram of the donor star in an initial binary system of 
a 0.7$M_{\odot}$ WD and a 2$M_{\odot}$ MS star and initial orbital period 
of 2.4 days. Blue indicates the phase of mass transfer onto the WD. The mass 
accretion rate is based on Hachisu et al. (1996, 1999). Mass transfer starts 
when the donor star is in the Hertzsprung gap and continues on the GB. After 
mass transfer the donor evolves into a He WD. The final system is a WD of 
$\sim$ 1.5 $M_{\odot}$ and a companion of $\sim$ 0.3 $M_{\odot}$. The crosses 
indicate different times after mass transfer has ceased (10$^{6}$, 10$^{7}$, 
and 10$^{9}$ years). After 10$^{6}$ yr: the donor will appear as 
a low-mass He--star, 10$^{7}$ yr: as a hot He WD, and 10$^{9}$ yr: as a cooler 
He WD. Di Stefano, Voss \& Claeys (2011).
(Courtesy of Rosanne Di Stefano @AAS. Reproduced with 
 permission.)}
\label{}
\end{figure}

\noindent{\it Faint white dwarf companions:} 

\noindent
During the spin--down stage, the companion star of the WD continues to evolve
and it could lose all its envelope (if it had not lost it at the end of the
spin--up stage already), its core becoming electron--degenerate and the 
star, then a second WD, starting to cool down and be ever dimmer. That is the
original spin--up/spin--down mechanism 
(Di Stefano, Voss \& Claeys 2011; see Figure 5). 

\smallskip
\noindent
If the spin--down stage lasted long enough, at the time of the SN explosion 
the ejected, surviving companion could be dim enough to escape detection 
in the searches made up to now in the central regions of the remnants of 
SNe Ia. It would also 
explain (Justham 2011) the absence of H lines in the nebular spectra of SN Ia 
(Leonard 2007).  

\smallskip
\noindent
There are several unknowns affecting the explicative power of the 
spin--up/spin--down mechanism. A first one, pointed out by Meng \& 
Podsiadlowski (2013) is whether a WD can actually gain significant angular 
momentum by accretion of material from its companion. Based on the evidence 
gathered from fast--spinning WDs in close binary systems, it may seem that
a strong magnetic field were necessary for spin--up, not just gas friction. 

\smallskip
\noindent
Even if the WD is spun--up significantly, solid--body rotation only slightly
increases $M_{\rm crit}$ above $M_{\rm Ch}$ (Yoon \& Langer 2004, 2005). 
Only differential rotation can make WDs with masses up to $\approx$ 2 
$M_{\odot}$, and even higher, stable. Therefore, the spin--down time scale
is set by that of the redistribution of angular momentum inside the WD, 
approaching solid--body rotation. Such time scale is very uncertain (and 
also the distribution of angular momentum at the start of spin--down): Yoon 
\& Langer (2005) find an upper limit of $\sim$ 10$^{6}$ yr only. Acknowledging
the difficulty of a theoretical determination, Meng \& Podsiadlowski (2013)
adopt a semi--empirical approach: assuming the companion to be a red giant 
at the end of spin--up, and based on the presence of circumstellar material 
coming from the envelope of the companion, in some SNe Ia at least, they
set an upper limit of $\sim$ 10$^{7}$ yr to the spin--down time scale. It
is, therefore, unclear that the companion stars could be dim enough, at the
time of the SN explosion, to easily scape detection.

\smallskip
\noindent 
A possible objection to the spin--up/spin--down mechanism is that the bulk 
of the SNe Ia are not super--Chandrasekhar: the Galactic SNe Ia whose light
curves have been reconstructed  (Ruiz--Lapuente 2004,
2017) appear to be completely ``normal''. Since there is nothing to make the
spin--up stop when the WD reaches the Chandrasekhar mass, it may seem that
the super--Chandrasekhar SNe Ia should then be more common. But only, as 
noted above, differential rotation can support WDs with masses significantly
higher than the Chandrasekhar mass. If the WD rotates almost as a solid body, 
the only slightly super--Chandrasekhar mass would not have a noticeable effect
in the SNe Ia observable characteristics.
 That would also explain  why there is not a wide span of masses 
of WDs above the Chandrasekhar mass that explode as SNe Ia.

\smallskip
\noindent 
A further point is that nobody has yet calculated the effects that the
impact of the SN ejecta would have on a WD companion, however dim it were
before, save in the case of extremely close ones, ejected when they were
starting to merge with the WD that explodes (Shen \& Schwab 2017).

\noindent{\it Subdwarf B star companions:}

\noindent 
Based on the common--envelope wind model for SN Ia (Meng \& Podsiadlowski
2017), Meng \& Li (2019) have very recently presented new evolutionary 
calculations for binaries initially made of a WD plus a main--sequence
companion. They find that, including the spin--up/spin--down mechanism, the
companion, at the time of the SN explosion, could either be a 
main--sequence star, a red giant, or a subdwarf B (sdB) star, that 
assuming that the spin--down time scale were $\lapprox$ 10$^{7}$ yr (the
upper limit derived in Meng \& Podsiadlowski 2013). 

\smallskip
\noindent 
In the common--envelope wind model, there is no merging of the WD with its
companion, the SN Ia taking place either in the common--envelope phase, in
a phase of stable H burning, or in a weakly unstable phase of H burning. The
companion type when the SN occurs depends on the initial parameters of the
system, the sdB type generally corresponding to the highest mass ratios 
$q \equiv M_{2}/M_{1}$. The effective temperatures and luminosities of the
sdB companions, at explosion, are in the range 30000 -- 40000 K and
10$L_{\odot}$ -- 65$L_{\odot}$, respectively. Meng \& Li (2019) estimate that
$\approx$ 22\% of the initial WD + main--sequence star systems should end
up as WD + sdB. Given the typical $T_{\rm eff}$ of these companions, they
should be searched, preferentially, in the U or UV bands.

\section{Remnants of Type Ia supernovae for the exploration of surviving
 companions}

\subsection{Strategy of the searches}

\noindent
The search for possible surviving companion stars of SNe Ia in SNRs
previously identified as produced by SNe of this type, starts with
the determination of the site of the explosion and of its uncertainty. A first 
approximation is the geometrical centroid of the SNR. If the ejection of 
material was spherically symmetric and the circumstellar and interstellar 
media homogeneous around the location of the SN, the edge of the SNR, 
projected on the sky, should appear circular and the approximation would be 
accurate. However, even in almost circular SNRs, some degree of asymmetry 
exists. By that reason alone, the region to be explored must cover, even 
in those cases, a significant fraction of the radius of the SNR. 

\smallskip
\noindent
Even if the explosion site were very accurately known, the companion must have 
left the binary system with the orbital velocity it had at the time of the 
explosion, plus the kick imparted by the collision with the SN ejecta. That 
should translate into proper motion (in any direction), so the star will be 
found at some angular distance from the explosion site. The highest velocities
expected, for still thermonuclearly evolving companions, are for main--sequence
stars, and they are adopted to estimate the maximum angular distance to be
covered, that depending, of course, on the age of the SNR. Possible surviving
WD companions could move much faster and the area to be covered when searching
for them increases accordingly (Kerzendorf et al. 2018b).

\smallskip
\noindent
The stars to be analyzed must not only be inside some area in the sky (size 
dependent on the above considerations), but also at distances compatible with 
that of the SNR. The latter  are not very accurately known in the case of most 
Galactic SNRs. In the case of the SNRs in the LMC, all stars within the 
searched area can be taken as being at the same distance, coinciding with that 
of the SNR. In the Galactic case, since the stars in the sample to be studied 
are physically unrelated, before the advent of the {\it Gaia} space mission  
the distances had to be determined star by star, based on comparison of their
spectra with photometric measurements.

\smallskip
\noindent
Spectra are needed in all cases to measure radial and rotational velocities, 
looking for kinematic peculiarity. Large telescopes are required for that. 
The 10m {\it Keck} and {\it Subaru} telescopes in Hawaii and the 4.2m William 
Herschel telescope in La Palma have been used in the case of SN 1572 (the 
only ``historical'' SN Ia that can be studied from the Northern Hemisphere). 
The 8.2m ESO VLT telescopes have dealt with the other two Galactic SNe Ia 
explored, SN 1006 and SN 1604, using multi--slit spectrographs. Only the {\it 
HST} can, at present, reach the sites of the SNe Ia in the LMC and this is 
for comparing the color and magnitude of possible companions with the 
expected values.

\smallskip
\noindent 
For the Galactic SN Ia potential surviving companions, radial velocities 
are directly measured from the spectra. 
With high--resolution spectra, the stellar atmosphere parameters of the 
surveyed stars are obtained by comparison with synthetic spectra.
The stellar parameters are required for 
distance determinations and are obtained by modeling of the stellar spectra.
Rotational velocities can be measured from high--resolution spectra as well.

\smallskip
\noindent
Another point concerning the detection of anomalously high space velocities
is what sets the comparison standards. The Besan\c con model of the Galaxy 
(Robin et al. 2003) has been used to that end in several studies. Another way 
to single out stars with peculiar motions is to use Toomre diagrams (plots 
of the velocities on the meridian plane of the Galaxy against those on the 
Galactic plane) in which the companion candidates are plotted together with
representative samples of the different Galactic populations (thin and thick
disc, bulge, halo). Here, the second data release of the 
 {\it Gaia} space mission ({\it Gaia} DR2) provides a new resource: 
to compare the velocities of the possible companions with the observed 
distribution of a large sample of stars at similar distances and positions
on the sky. The scrunity of kinematics of the stars through the {\it Gaia} DR2
is limited to distances of 2--3 kpc away from us.

\smallskip
\noindent  
The {\it Gaia} DR2 makes also posible to measure trigonometric 
parallaxes and 
proper motions for candidates to supernova companion, in the SNRs of the 
Galaxy, that  are brighter than $G \simeq$ 20--21 mag ({\it Gaia} 
white--light magnitudes). 
In particular, several studies have been made in some of the 
SNe Ia that we will present. The proper motions of {\it Gaia} are given in 
an absolute frame (the ICRS), whereas those from {\it HST} are always
 relative to a local frame. Comparison is possible (see for instance in Tycho;
 Ruiz--Lapuente et al. 2019). 

\smallskip
\noindent
All the precedent is meant for searches in SNRs that come from a thermonuclear
explosion. Therefore, the first step in any search is to identify the SNR as a 
SNR coming either from a thermonuclear explosion or from a core collapse supernova. 

\smallskip

\subsection{Typing a SN Ia SNR}

\smallskip

As already said, there are about 300 SN remnants in our
 Galaxy  (Green 2014) and an undertermined number in
the LMC. The number of unambiguosly  classified remnants
(thermonuclear, SN Ia SNRs and core collapse CC, SN II, Ib SNRs), however, is
much smaller. 
One early way of classifying them as SNe Ia remnants or as core collapse 
remnants was through their morphology. Highly asymmetric X--ray morphology is
typical of core collapse remnants (Lopez et al 2011). There is, however, a 
much better classification which emerges 
from the Fe--K shell X--ray emission (6--7 keV
band), as explained in Yamaguchi et al. (2014).  

\noindent
The Fe--K$\alpha$ centroid is in the red part of the spectrum in SN Ia 
remnants and in the blue part in CC SNe (see Yamaguchi et al. 2014;
 Martinez--Rodriguez et al. 2017). To discriminate explosion properties 
amongst SNe Ia models, line flux ratios (Si K$\alpha$/Fe K$\alpha$, 
S K$\alpha$/Fe K$\alpha$, Ar K$\alpha$/Fe K$\alpha$) are useful.
It is even possible to discriminate  the metallicity of the SN Ia 
remnants from other ratios involving the Cr and Mg abundances 
(Badenes et al. 2008).   

\noindent
The three youngest and closest  Galactic SNRs of SN Ia (1006, Tycho and 
Kepler) have been studied very intensively 
in the context of possible stellar companions. We also 
include in Table 3 all SNe Ia with good 
X--ray data that have been classified as SNe Ia in our Galaxy and in the LMC.
We will discuss them briefly. We start giving an extensive account of the
three first mentioned SN Ia remnants.

\begin{table*}
\begin{center}
%\footnotesize
\caption{Distances, size and ages of the Ia SNRs
 \label{table:remmants}}
\begin{tabular}{cccc}
\hline
\noalign{\smallskip}
\hline
\noalign{\smallskip}
Name
& Distance & Size (Radius) & Age \\
\noalign{\smallskip}
 & ($\mathrm{kpc}$)  & ($\mathrm{arcmin}$) & ($\mathrm{years}$)  \\
\noalign{\smallskip}
\hline
\noalign{\smallskip}
			
Kepler & $ 5 \pm 0.7 $ & $ 1.9 $ & $414$  \\
			
3C 397 & $6.3-9.8$ & $1.7$ & $1350-1750$   \\
			
Tycho & $2.83 \pm 0.79 $ & $ 4.3 $ & $446$   \\
			
RCW 86 & $2.5$ & $21$ & $1833$   \\
			
SN 1006 & $2.18 \pm 0.08 $ & $15 $ & $1012$ \\

G1.9+0.3 & $8.5$  & $0.8$ & ${\sim} 150$  \\

G272.2-3.2 & $1-3.2$ & $10$ & $8000$  \\
			
G337.2$-$0.7 & $2.0-9.3$ & $ 3 $ & $ \sim 5000 $ \\

G299.2-2.9 & $5$ & $5$ & $5000$   \\
			
G344.7$-$0.1 & $6-14$ & $5$ & $3000-6000$  \\
			
G352.7$-$0.1 & $7.5$ & $3.5$ & ${\sim} \, 2200$   \\
			
N103B & 50 & $0.2$ & ${\sim} \, 860$  \\
			
0509$-$67.5 & 50 & $0.27$ & ${\sim} \, 400$  \\
			
0519$-$69.0 &  50  & $0.3$ & ${\sim} \, 600$   \\

0548$-$70.4 & 50 & $0.9$ & $10,000$ \\  
			
DEM L71 & 50 & $0.7$  & ${\sim} \, 4700$  \\

\noalign{\smallskip}
\hline
\end{tabular}
\end{center}
\noindent
{\bf References:}  Ruiz--Lapuente (2017); Yamaguchi, H., et al. (2015); Ruiz--Lapuente (2004); Bocchino et al. (2000); Winkler et al. (2014); Borkowski et al. (2017); McEntaffer et al. (2013); Rakowski et al. (2003); Post et al. (2014); Giacani et al. (2011); Sezer \& 
G\"ok (2014); Sano et al. (2018); Litke, Chu \& Holmes (2017); Edwards, Pagnotta \& Schaefer (2012); Hendrick, Borkowski \& Reynolds (2003); Hughes, Hayashi \& Koyama (1998); Kinugasa et al. (1998). 
\end{table*}

\begin{figure}
\includegraphics[width=0.75\columnwidth, angle=0]{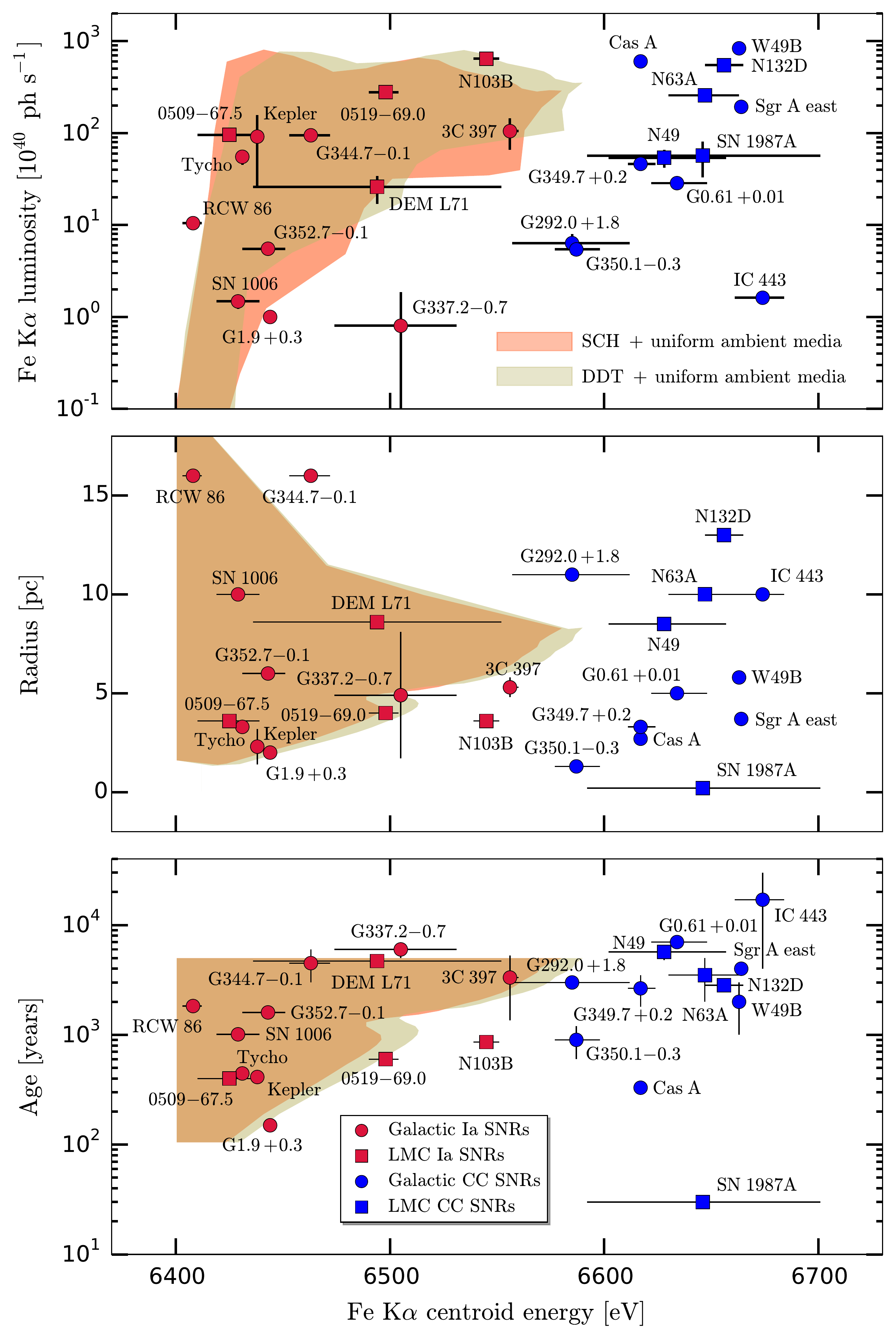}
\caption{ Fe K$\alpha$ luminosity, radius and expansion age as a function of the Fe K$\alpha$ centroid energy for Ia (red) and CC
(blue) SNRs. The shaded regions depict
the predictions from the theoretical MCh (khaki) and sub-MCh (dark orange) models with uniform ISM densities of 
Mart\'{\i}nez--Rodr\'{\i}guez et al. (2018). 
(Courtesy of Hector Mart\'{\i}nez--Rodr\'{\i}guez and Carles Badenes. @AAS. 
Reproduced with permission).}
\label{Figure 1}
\end{figure}

\subsection{SN 1572 (Tycho Brahe's supernova)}

The central region of the remnant of the SN that appeared in 1572 (also 
known as Tycho Brahe's supernova) was the first to be explored in 
search of a possible surviving stellar companion of the SN (Ruiz--Lapuente et 
al. 2004). The 2.5m {\it Isaac Newton Telescope} (photometry), the 4.2m 
{\it William Herschel Telescope}  (spectroscopy), and the 2.5m {\it Nordic 
Optical Telescope} (spectroscopy), at the observatories in the Canary Islands, 
were mainly used, as well as the {\it HST} (astrometry). 
Supplementary observations were made with the two 10m {\it Keck} telescopes 
(spectroscopy), in Hawaii.

\smallskip
\noindent
The region searched was a circle of 39 arcsec radius (about 15\% of 
the radius of the SNR), with the same center as the X--ray emission measured 
by the {\it Chandra} satellite (RA = 00 h, 25 min, 19.9 s; dec = 64$^{o}$, 
08', 18.2''). The limiting apparent visual magnitude of the survey was $V$ = 
22. The distance to the SNR being $\sim$3 kpc (2.83$\pm$ 0.79 kpc in 
Ruiz--Lapuente 2004), and the visual extinction 
$A_{V} = 1.7-2.0$ mag, all main--sequence stars of spectral type earlier than 
{\it K6} must have been detected (the Sun would appear, from there, as a
$V = 18.9$ mag star). 
       
\smallskip
\noindent
The distances to the stars were determined from their spectral types and
luminosity classes as compared with their visual magnitudes, taking into 
account extinction. Radial velocities were measured from the spectra 
and proper motions from the {\it HST} images (using the {\it WFPC2}). No 
chemical 
abundance analysis was attempted for any of the observed stars, just the 
metallicity being estimated. No rotational velocities were obtained, 
either.  

\smallskip
\noindent
Of the stars at distances compatible with Tycho's SNR,  
one star, labeled $G$ (located at 29.7 arcsec to the SE of the adopted center 
of the SNR) had an unusually high radial velocity (later refined to $v_{r} = 
-80 \pm 0.5$ km s$^{-1}$, by Gonz\'alez Hern\'andez et al. 2009, or to $-79 
\pm 2$ km s$^{-1}$, by Kerzendorf et al. 2009, both velocities in the LSR), to 
be compared with an average velocity $v_{r} = -20$ to $-40$ km s$^{-1}$, with 
a $\sim$20 km s$^{-1}$ dispersion, at the distance and position of the SNR. 
The proper motions were found to be $\mu_{b} = -6.11 \pm 1.34$ mas yr$^{-1}$, 
perpendicularly to the Galactic plane, and $\mu_{l} = -2.6 \pm 1.34$ mas 
yr$^{-1}$, parallel to it. At the distance of the SNR, that would 
mean a tangential velocity of $94 \pm 27$ km s$^{-1}$.
 The modulus of the total velocity vector (radial 
plus tangential), would be 136 km s$^{-1}$, a factor of 3 larger than the mean 
velocity  at 3 kpc, according to approximative estimates.    

\smallskip
\noindent
The atmosphere parameters of star $G$ were $T_{eff} = 5,750$K, log $g$ between
3 and 4, and metallicity close to solar (its lower limit was [M/H] $> -0.5$). 
That corresponded to a G0--G2 subgiant of mass about 1 $M_{\odot}$ and radius 
$R = 1-3\ R_{\odot}$. The star, before the explosion, might either have been a 
main--sequence star, now still puffed--up after the SN impact, or a subgiant 
with a larger mass. 

\smallskip
\noindent
Based on its kinematic peculiarity, its distance and its profile, 
Ruiz--Lapuente et al. (2004) pointed to star $G$ as a likely candidate to
have been the companion of SN 1572.

\smallskip
\noindent
Such identification has later been disputed. Kerzendorf et al. (2009) 
objected that high rotational velocities must be a characteristic of the
companions of SNe Ia. That was based on the fact that the companion should
be co--rotating with the system, at the time of the explosion (that is, the
rotation period being equal to the orbital period). The orbital period being 
short in a contact binary, one of its components being a white dwarf and the
other a main--sequence star filling its Roche lobe, the companion would be a 
fast rotator, but star G rotates slowly: $v_{rot}$ sin $i \lapprox$ 6.6 km 
s$^{-1}$ (Gonz\'alez Hern\'andez et al. 2009) or $v_{rot}$ sin $i \lapprox$  
6 km s$^{-1}$ (Kerzendorf et al. 2013).  In the Section on rotation above, 
however, we have discussed the reduction of the pre--explosion rotational 
velocity by the interaction with the SN ejecta. Based on that, Pan, Ricker \& 
Taam (2012b) and Liu et al. (2013) argue that the slow rotation of star G 
does not discard it as a possible companion of the SN. However,  
Liu et al. (2013) find that, given the age of Tycho's SNR, the rotational 
velocity of their most slowly rotating model ($\sim$ 25 km/s) after impact
is still higher than the measured rotational velocity of star G.
The models by Pan, 
Ricker \& Taam (2012b; their Table 3) with low rotation 
 are of bloated stars. One can argue then that the models do 
not fully reproduce the characteristics of Tycho G.

\smallskip
\noindent
Gonz\'alez Hern\'andez et al. (2009), from a high--resolution 
{\it Keck-1/HIRES} spectrum refined the stellar parameters of star G:
$T_{\rm eff}$ = 5900$\pm$100 K, log $g$ = 3.85$\pm$0.30, [Fe/H] = 
-0.05$\pm$0.09. Also, a Ni overabundance relative to Fe, 
[Ni/Fe] = 0.16 $\pm$ 0.04 was measured, about 3$\sigma$ above the average 
value in Galactic disk stars (Ecuvillon et al. 2004,2006; Gilli et al. 2006), 
which they attributed to pollution by the SN ejecta. 

\smallskip
\noindent
Kerzendorf et al. (2013), however, from the same {\it Keck} spectrum, derived
[Ni/Fe] = 0.07 $\pm$ 0.04 only. They attributed the difference to differences
in equivalent width (EW) measurements of Ni lines, maybe related to continuum 
normalization and/or local continuum placement. They compared the ratio to 
that from the set of F-- and G--dwarf abundances of Bensby et al. (2005) and 
found that the value was not unusually high. From that, Kerzendorf et al. 
(2013) concluded that star G was likely a backgroud star, unrelated to
the SNR.     

\smallskip
\noindent
Bedin et al. (2014) have later redetermined the Ni abundance (again from the
same spectrum), using automated tools to fit the continuum and measure the 
EWs of the Ni lines. They have also removed weak lines. They find 
[Ni/Fe] = 0.10 $\pm$ 0.05. Thus, the three measurements are compatible with 
each other, within the errors. The latter abundance ratio, when compared with 
the Galactic trend from Neves et al. (2009) (different from that in Bensby 
et al. 2006), still appears to be almost 1.7$\sigma$ above the trend. From 
that, these authors conclude that the probability of the combination of 
the peculiar motion with even that moderate overabundance, in a star unrelated 
to the SN, would still be very low. Also, in Bedin et al. (2014), accurate 
proper motions were determined, from {\it HST} astrometry, for 1148 stars in 
the central region of Tycho's SNR. Proper motions for 16 of these stars had 
also been measured by Kerzendorf et al. (2013), with good agreement for the 
stars common to the two sets.

%\smallskip
%\noindent
%Concerning proper motion, there is agreement between Kerzendorf et al. (2013)
%and Bedin et al. (2014): $\mu_{\alpha} = -2.50 \pm 0.60$ mas yr$^{-1}$ and 
%$\mu_{\delta} = -4.22 \pm 0.60$ mas yr$^{-1}$ (Kerzendorf et al. 2013); 
%$\mu_{\alpha} = -2.63 \pm 0.19$ mas yr$^{-1}$ and 
%$\mu_{\delta} = -3.98 \pm 0.11$ mas yr$^{-1}$ (Bedin et al. 2014).      

%\begin{figure}
%\includegraphics[width=1.0\columnwidth, angle=0]{f5_color.eps}
%\caption{Two portions of the spectrum of star G, showing lines of Ni, Fe, and 
%Ca (from Gonz\'alez Hern\'andez et al. 2009).}
%\label{Figure 1}
%\end{figure} 

\smallskip
\noindent
Star $G$ is not the only star in the central region of Tycho's SNR to have
been proposed as a possible companion of the SN. Ihara et al. (2007) 
claimed that the spectrum of star $E$ showed blueshifted Fe absorption lines, 
that they interpreted as due to absorption by the approaching part of the 
expanding SNR. Star E, however, from the parallax
measured by the space mission {\it Gaia} data release DR2, is found to be 
far behind the SNR (though the parallax determination has a large errorbar) .

\smallskip
\noindent
Kerzendorf et al.(2013) had pointed to star $B$. It is a 
peculiar A--star, exhibiting fast rotation and an unusual abundance pattern of 
low overall metallicity, [Fe/H] = -1.1, yet high abundances of C and O. It is 
a few arcsecs from the geometrical center of the SNR, and at an estimated 
distance consistent with that of the remnant. Its surface temperature is 
$T_{\rm eff}$ = 10,722 K, log $g$ = 4.13, and $v_{rot} = 170$ km s$^{-1}$. The 
high rotational velocity, however, is not unusual for an A--star.
The abundance pattern can be explained if  Tycho B is, as it seems,  a 
$\lambda$ Bootis star (see, for instance, Paunzen 2004, for a review of their 
properties). The deficit in Fe--peak elements in this type of stars has been 
attributed to dust--grain formation by the more refractory elements and their 
ejection by radiation pressure, in an accretion disk around the star. In 
this way, the atmosphere of the $\lambda$ Bootis stars would be depleted of 
Fe--peak elements while having solar or near--solar abundances of C, N, O and 
S. The presence of such accretion disk would indicate that the now single star
was a member of a binary system. However, star B as candidate
 companion of SN 1572
does not fit the characteristics of an impacted star that would be 
contaminated by iron--peak elements and would have a kinematical imprint
that is not found in the measured proper motions and 
radial velocities of star B.

\smallskip
\noindent
More recently, Kerzendorf et al. (2018a) have obtained UV spectra of star B 
with the STIS low--resolution grating of the {\it HST}, in search of broad 
Fe II absorption features due to the SNR, which would have shown that the star 
was inside or behind the remnant. From their absence and a new luminosity
distance to star B, Kerzendorf et 
al. (2018a) conclude that it is a foreground star.  However, the data of 
the {\it Gaia} DR2 place it at a distance compatible with that of the Tycho 
SN, but the star, as already said, does not present any kinematic peculiarity. 

\smallskip
\noindent
There had already been restrictions set on the high energy emission of the
progenitor systems of the SN Ia, based on the state of the interstellar medium 
of their 
host galaxies (Woods \& Gilfanov 2013, 2014). If, in the single--degenerate 
channel for production of the SNe Ia, mass accretion by the WD that will 
eventually explode is mediated by thermonuclear burning, at its surface, of 
the material transferred from the companion star, that would generate luminous 
line emission which should be seen when observing the host galaxies if the 
rate of production of SNe Ia through the SD channel were high. 
From that, Johansson et al. (2014, 2016) have concluded that this accretion 
model could only contribute by a few percent to the total SN Ia rate in 
passively 
evolving galaxies, where there are no longer hot stars feeding such emission.   
Woods et al. (2018) have now set restrictions to the temperatures and 
luminosities of the progenitors of individual SNRs of the Ia type, based on 
such expected line emission luminosity. In the case of Tycho's SN,  
the SD scenario is also disfavored on these grounds (Woods et al. 2017).

\smallskip
\noindent
Another point, concerning not only the case of Tycho but all the SNRs that 
have been or will be surveyed in search of surviving companions of SN Ia,
is the exact location of the site of the explosion. A centroid of the 
remnant can always be found and serve as initial guide for the exploration, 
but there can be sizeable shifts between the centroid and the actual site. 
Even for a perfectly symmetric explosion, if the SN ejecta encounter a density 
gradient in the circumstellar or the interstellar medium, in some direction, 
the expansion will be slowed down as it escalades the positive gradient, and 
accelerated as it runs down the negative one, so the apparent center of the 
resulting SNR will be shifted towards the side of the decreasing densities. 
That has been shown by the hydrodynamical simulations of Williams et al. 
(2013), where the remnant keeps a round shape in spite of this asymmetry.

\smallskip
\noindent
On the other hand, there can be initial asymmetry in the SN ejecta themselves, 
as illustrated by Winkler et al. (2005) in the case of SN 1006. The 
structure inferred from absorption observations of background objects gives
an explosion center displaced from the geometrical center of the SNR by an
angular distance $\approx$ 19\% of the remnant's radius. In the case of 
Tycho, Krause et al. (2008), from the spectrum of the light echo of the SN, 
also suggest that the explosion was aspherical.

\smallskip
\noindent
Xue \& Schaefer (2015), from a combination of historical reconstruction and
semianalytical, approximative hydrodynamics, place the explosion site of
SN 1572 at 37 arcsec ($\approx$ 15\% of the SNR radius) to the NW of the 
geometrical center of the SNR. But more recently, Williams et al. (2016), 
by combining new measurements of the proper motions of the forward shock 
of the expanding material 
with hydrodynamical simulations, determine a site located 
22.6 arcsec ($\approx$ 9.5\% of the radius) to the NE of the geometrical 
center. This is more consistent with the existence of a density gradient
in the E--W direction in the interstellar medium, with the density increasing 
towards the E. The hydrodynamical simulations of Williams et al. (2016) 
assume spherical symmetry in the ejecta which, as we have seen, could not be 
true, that leaving an uncertainty range that does not allow to exclude any of 
the stars in the surveyed area.

\smallskip
\noindent
Very recently, Ruiz--Lapuente et al. (2019) have used the data release DR2 
of the {\it Gaia} space mission to reevaluate distances and proper motions
of the stars in the Tycho field.  They have looked at the stars within 1 degree 
from the centroid of the SNR, and at a distance 
compatible with it. They examine the stars comprised in
 an area four times larger than in the work in 2004.
 They find general agreement with the
distances estimated previously, with some exceptions, though. Star G is
found to be somewhat closer than in earlier measurements. A Toomre diagram 
(see Figure 7) 
shows that its kinematics is similar to that of thick disk stars, its
chemical composition being typical of the thin disk, however. Only $\approx$
0.8\% of stars share these characteristics. The orbits
described in the Galaxy by representative stars of the sample and by stars G 
and U are calculated and compared. These two last stars reach, by far, the 
highest distances from the Galactic plane, the total velocity of star U
being significantly smaller. The very large number of stars with precise
distances and proper motions in the {\it Gaia} DR2 now allow to compare
the proper motions of candidate stars with those of a huge sample of stars
around the same position and within the same range of distances (see 
Figure 8). Only stars G and U lie more than 2$\sigma$ above average.
These authors conclude that if star G were not the SN companion, the DD
channel should be preferred for the origin of Tycho's SN.

\begin{figure}
\includegraphics[width=0.8\columnwidth, angle=0]{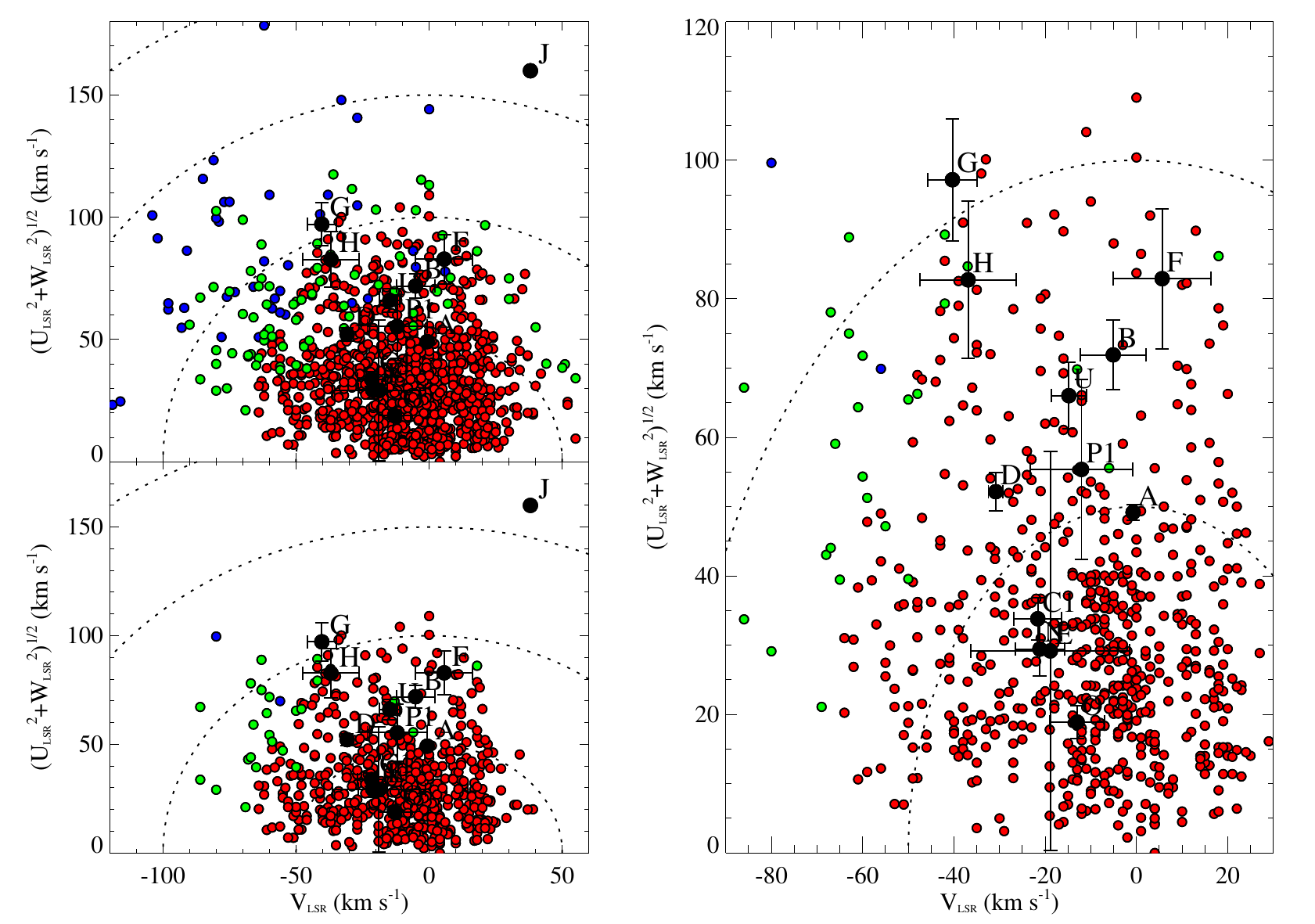}
\caption{Left upper panel: Toomre diagram for a sample of thin 
disk, thick disk, and transition thin--thick disk stars, covering a wide range 
of metallicities, with stars in the Tycho field 
 superimposed (red dots correspond
to thin disk stars, green to transition, and blue to thick disk 
stars). Left lower panel: same as upper panel, keeping only stars with
metallicities equal to or higher than that of star G. Right panel: detail 
of the lower left panel, leaving out star J, due to its large uncertainties 
in parallax and  proper motions. The sample is taken from
Adibekyan et al. (2012). (Ruiz--Lapuente et al. 2019. @AAS. 
Reproduced with permission).}
\label{Figure 3}
\end{figure}

\begin{figure}
\includegraphics[width=0.8\columnwidth, angle=0]{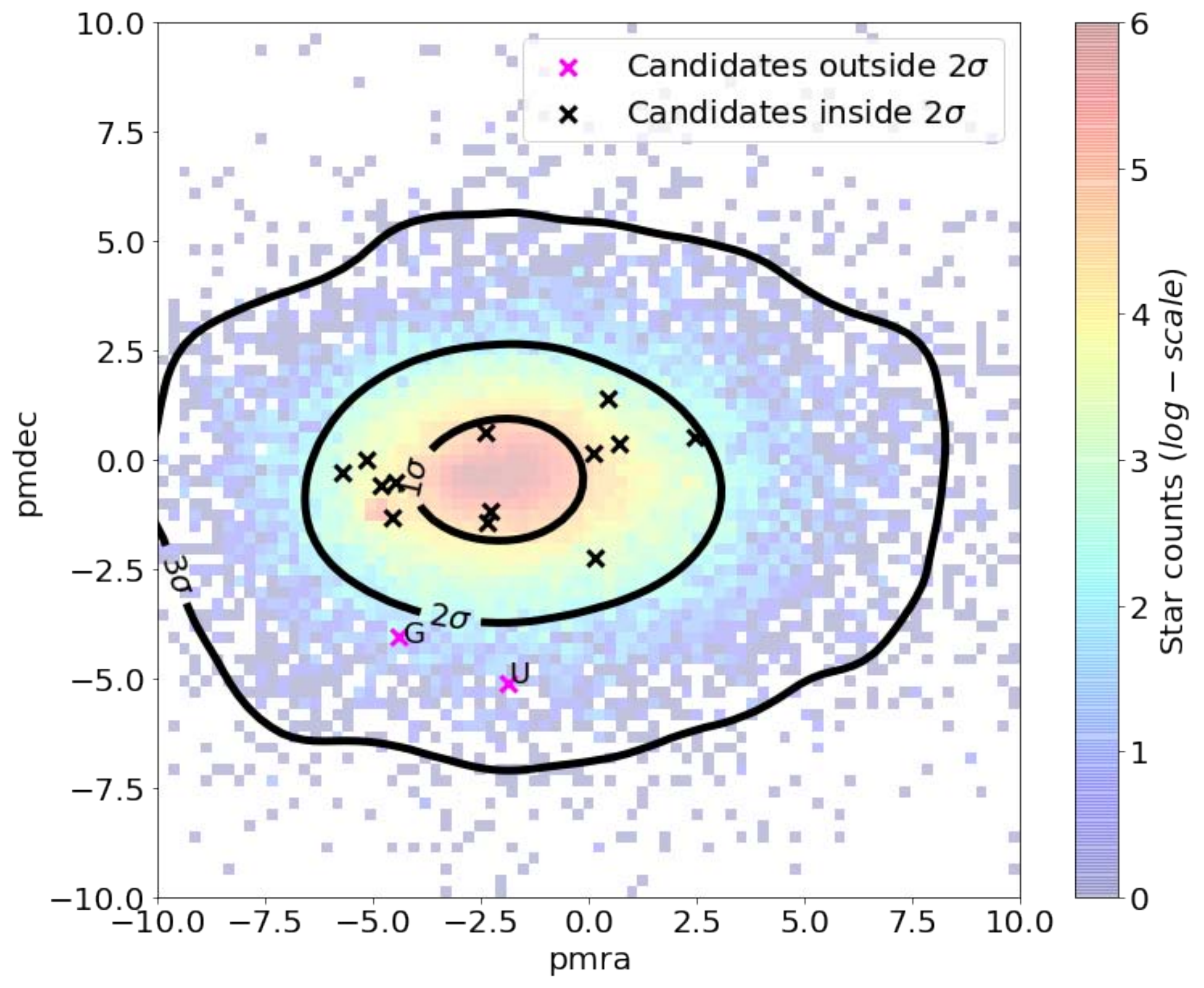}
\caption{Proper motions of stars G and U, superimposed on those of a 
very large sample of stars around the same position and within the same
range of distances. (Ruiz--Lapuente et al. 2019. @AAS. 
Reproduced with permission).}
\label{Figure 2}
\end{figure}

\begin{figure}
\includegraphics[width=0.8\columnwidth, angle=0]{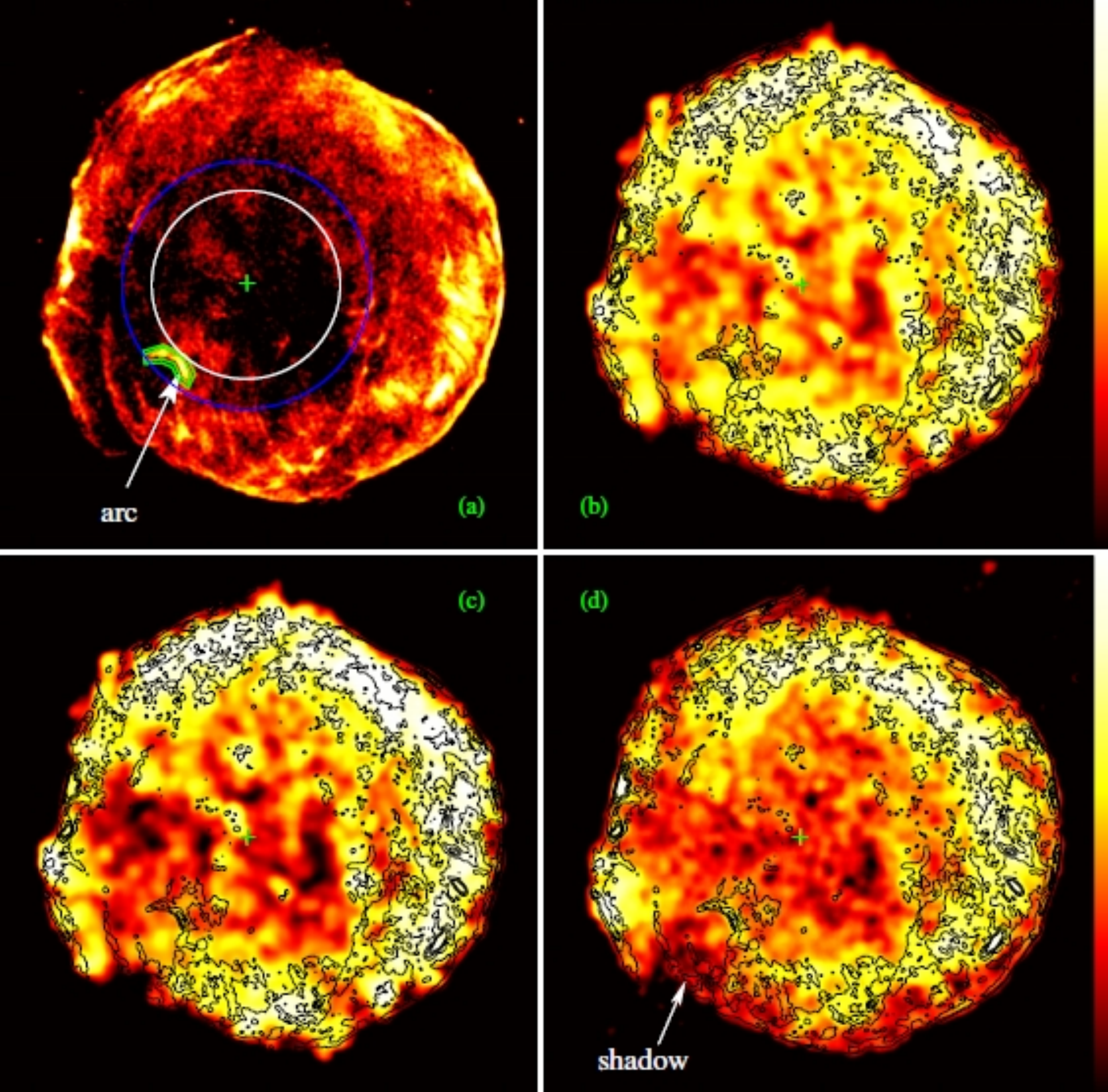}
\caption{Chandra images of the Tycho SN remnant in four different energy bands.
In the first panel, the arc attributted to the interation of the SN ejecta 
with the companion is indicated. In the fourth panel the ``shadow'' in the 
emission that would be produced by the companion is also indicated. (Lu et.
al. 2011). (Courtesy of Fangjun Lu @AAS. 
Reproduced with permission).}
\label{Figure 2}
\end{figure}

\smallskip
\noindent
Lu et al. (2011)  find a nonthermal X--ray feature 
in the SNR that  seems to result
from interaction between the SN ejecta and the stripped mass of the companion,
aligned with the radial direction of Tycho G. They suggest this as a  
property favoring a SD origin of this SN.
In this paper (Figure 9),  and also as predicted 
in Marietta, Burrows \& Fryxell (2000) and in Garc\'{\i}a--Senz et al. 
(2012), the supernova ejecta intercepted by the companion leave a 
''shadow'' behind that is visible in X--rays. 
 Zhou et al. (2016), from radio observations, find that Tycho is
surrounded by a clumpy, expanding molecular bubble, whose origin would be a
fast outflow driven from the vicinity of a WD as it accreted matter from a
nondegenerate companion star.
 
\smallskip
\noindent
The SD scenario for this supernova is still debated, however

\subsection{SN 1006}

SN 1006 has always been classified as a Type Ia supernova (see, for instance, 
Stephenson 2010). Its position on the sky ($\delta_{\rm J2000} \simeq$ 
-42$^{o}$) makes any deep study of its remnant only suitable for telescopes in 
the Southern Hemisphere.  

\smallskip
\noindent
The distance to the SNR has been determined, from the expansion velocity
and the proper motion of the ejecta, to be $d = 2.18 \pm 0.08$ kpc (Winkler, 
Gupta \& Long 2003). It is located about 500 pc above the Galactic plane. 
The interstellar extinction in the V--band is $A_{V} = 0.3$ mag only, much 
lower than in front of the two other historical Galactic remnants of SNe Ia, 
Tycho's and Kepler SNRs. 

\smallskip
\noindent
The central region of the remnant of SN 1006 (see Figure 10) 
 has been independently explored,
in search for a possible surviving companion, by Gonz\'alez Hern\'andez et al. 
(2012) (GH12, from now on) and by Kerzendorf et al. (2012) (K12 henceforth). 

\begin{figure}
\centering
\includegraphics[width=0.8\columnwidth, angle=0]{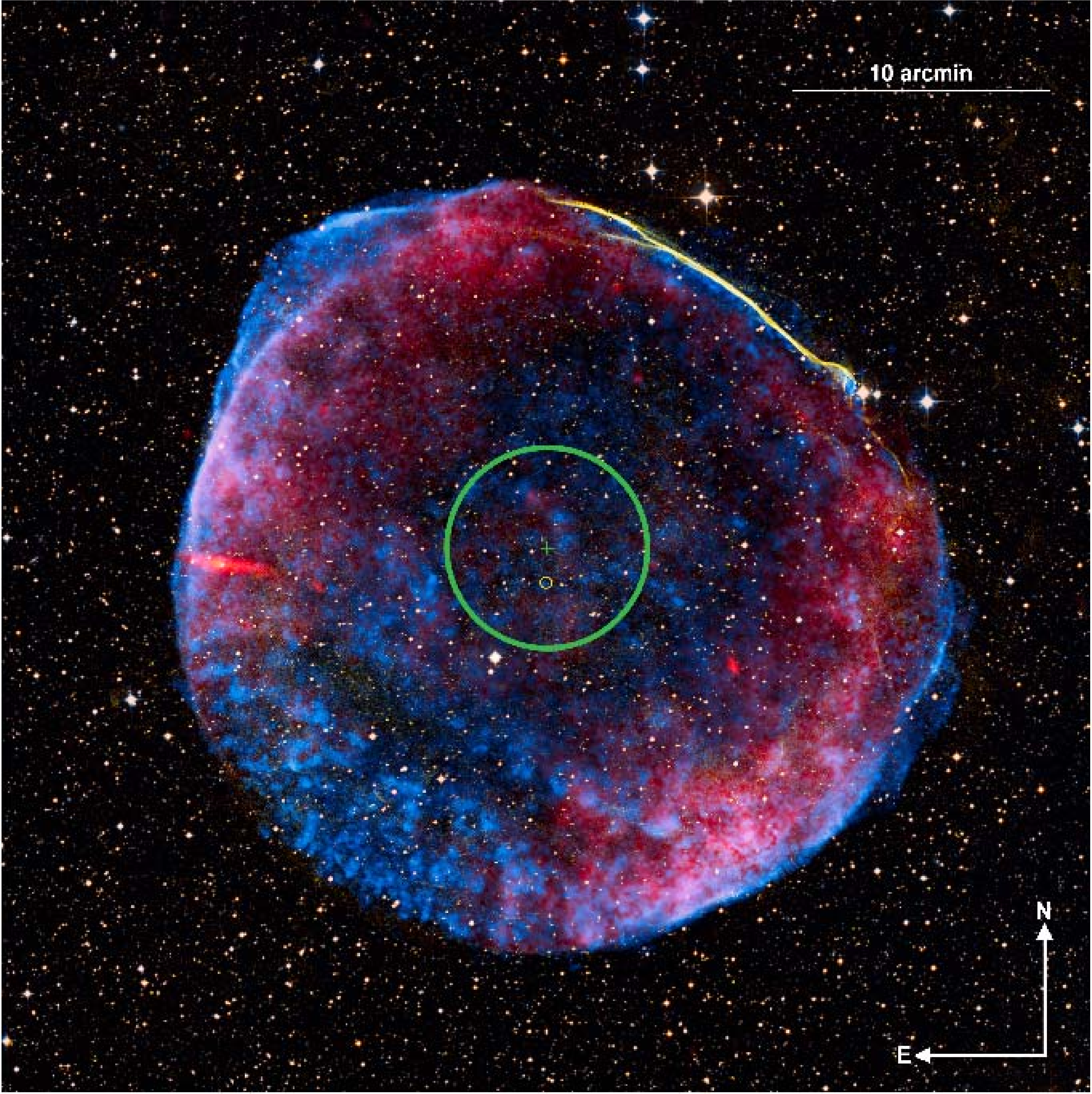}
\caption{The remnant of SN 1006. The surveyed area in GH12 is indicated by the
large green circle. The center of the survey (the centroid of the X--ray 
emission) is marked with a green cross, and that of the H$\alpha$ emission
by the small yellow circle. The image is a composite of the X--ray, optical,
and radio emissions (from GH12). (Gonz\'alez Hern\'andez et al. 2012.
 @Springer Nature. Reproduced with permission).}
\label{Figure 1}
\end{figure} 

\smallskip
\noindent
The search by GH12 covered a circle of 4 arcmin radius around the geometrical 
center of the X--ray emission of the remnant determined by Allen, Petre \& 
Gotthelf (2001): $\alpha_{\rm J200}$ = 15$^{h}$ 2$^{m}$ 55$^{s}$, 
$\delta_{\rm J2000}$ = -41$^{o}$ 55' 12''. That radius amounts to 27\% of the 
radius of the SNR (15 arcmin). K12 chose instead, for the center, a mean of 
the X--ray and radio centers, at $\alpha_{\rm J2000}$ = 15$^{h}$ 2$^{m}$ 
22$^{s}$.1, $\delta_{\rm J2000}$ = -41$^{o}$ 55' 49'', and a search radius of 
2 arcmin. We must note, however, that Winkler et al. (2005), based on the 
distribution of the ejecta along the line of sight, propose a different center 
of the explosion, very close to the position of the Schweizer--Middleditch 
star, at $\alpha_{\rm J2000}$ = 15$^{h}$ 2$^{m}$ 53$^{s}$.1, $\delta_{\rm J2000}$ 
= -41$^{o}$ 59' 16''.7. That is still inside the search radius of GH12, 
although close to the edge of the explored region.

\smallskip
\noindent
The limiting magnitude of the GH12 survey, in the R--band, was $m_{R} = 15$ 
mag. Given the distance and the extinction, that included all red giants, 
subgiants and main--sequence stars down to $M_{R} = +3.1$ mag, but from the
{\it Two Micron All--Sky Survey} ($2MASS$: Cutri et al. 2003) there are no 
main--sequence stars brighter than $m_{R} = 16.4$ mag in the explored field. 
That brings the limit down to $M_{R} = +4.5$ mag or, in the V--band, to 
$M_{V} = +4.9$ mag (stars only slightly less luminous than the Sun). 

\smallskip
\noindent
The K12 survey reached a limiting magnitude of $m_{V} = 17.5$ mag where they 
did full stellar modeling and a depth of $m_{V} = 19$ of stars for which
 they only 
probed the radial velocity. 
Given the distance and the extinction, $M_{V} = +5.5$ mag, and those limits are 
equivalent to half the Solar luminosity and a tenth of the solar luminosity,  
respectively.
 Thus, the GH12 survey was more extended than K12 
(four times in the area covered) but the latter was deeper. 

\smallskip
\noindent
Both GH12 and K12 surveys were spectroscopic, although K12 had, in addition, 
done a previous run of photometric observations. In the two cases the 
{\it Very Large Telescope} of the {\it European Southern Observatory} was 
used, but with two different instruments: the high--resolution {\it 
Ultraviolet and Visual Echelle Spectrograph} ($UVES$) in GH12, and the 
{\it Fibre Large Array Multi Element Spectrograph} ($FLAMES$) with the 
medium--high resolution $GIRAFFE$ spectrograph in K12.    

\smallskip
\noindent
In the two surveys, the stellar parameters $T_{\rm eff}$, log $g$, and the 
metallicity [Fe/H] were derived from the observed spectra, using similar
techniques. In addition, GH12 also determined the chemical abundances of the
Fe--peak elements Cr, Mn, Co and Ni, as well as those of Na and of the 
$\alpha$--elements Mg, Si, Ca and Ti. The spectra equally provided the 
radial and rotational velocities  of the stars.

\smallskip
\noindent
Thanks to the high quality of the UVES spectra, the errors in the stellar
parameters of GH12 are very small: from 30 to 100 K in $T_{\rm eff}$, from 
0.1 to 0.2 in log $g$, and from 0.03 to 0.06 dex in [Fe/H]. The
 FLAMES/GIRAFFE spectra, instead, make the errors in K12 
to be larger: 250 K in $T_{\rm eff}$, 0.5 in log $g$, and 0.5 dex in [Fe/H].    

\smallskip
\noindent
No significant rotational velocities were found for any of the stars, neither 
in the GH12 nor in the K12 surveys.

\smallskip
\noindent
As for the radial velocities, GH12 compared the observed velocities with the 
distribution predicted by the Besan\c con model of the Galaxy (Robin et al. 
2003): all the stars in the sample were consistent with the model 
distribution, with no significant outlayer being found.
GH12 compared the observed distribution of the abundances of Fe--peak 
elements as a function of metallicity with the Galactic trends (see 
Figure 11), in search for signs of chemical contamination of the surface of 
some star by the SN ejecta. All stars are within the dispersion of the 
Galactic trends. The same is true for the $\alpha$--elements. 
  
\begin{figure}
\centering
\includegraphics[width=0.9\columnwidth, angle=0]{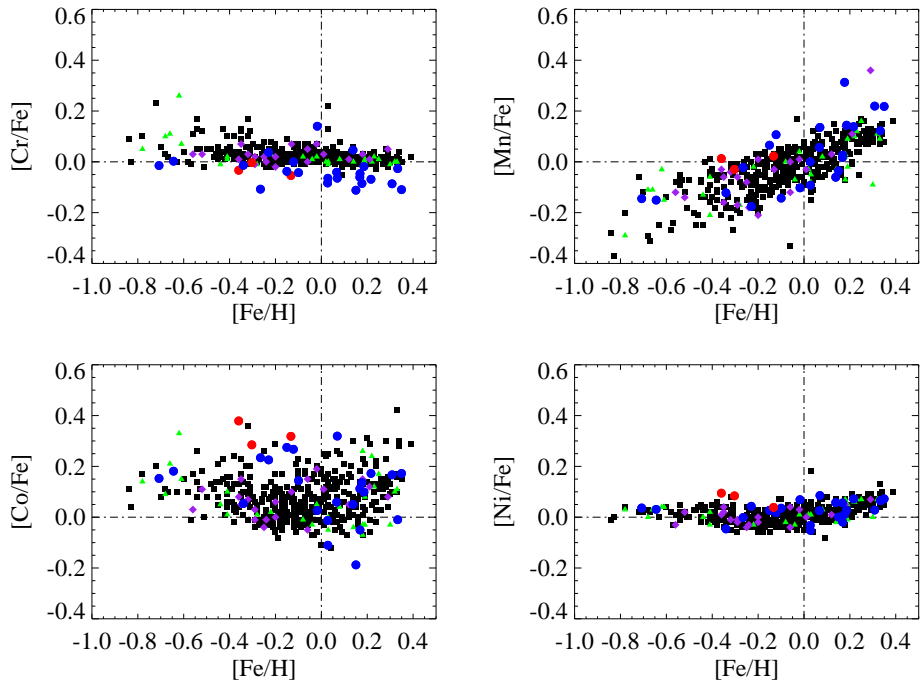}
\caption{Stellar abundance ratios [X/Fe] of several Fe--peak elements. Red 
triangles correspond to the four giant stars at distances compatible with
that of the remanant of SN 1006. Blue squares, to the rest of the stars in the
sample of GH12. (Gonz\'alez Hern\'andez et al. 2012. @Springer Nature. 
Reproduced with permission).}
\label{Figure 8} 
\end{figure}

\smallskip
\noindent
GH12 also use the stellar parameters derived for the sample stars, and the 
photometric magnitudes in five different filters from the {\it 2MASS} 
catalogue, to determine the distances. Only four stars are at distances
(marginally) compatible with that of the SNR. All of them are red giants, 
without any kinematic nor spectroscopic peculiarity.

\smallskip
\noindent
The conclusions of GH12 and K12 were the same: either the companion 
must have been an unevolved star, less luminous than the Sun, and 
having returned to its initial state only $\sim$1,000 yr after the 
SN explosion (which appears very unlikely) or the explosion was 
due to the merging of two white dwarfs. In both papers, the possibility
that the SN were produced through the spin--up/spin--down channel
and then the companion would be a faint WD at the time of the explosion,  
was considered, arguments against it being given in GH12 and in 
subsection 2.4 of the present paper. 

\noindent
The spin--up/spin--down channel is not the only way to have a WD as a 
surviving companion of a SN Ia. As discussed above, Shen \& Schwab (2017) 
have considered the
case of He detonations close to the surface of a mass--accreting C+O WD, 
induced by mass transfer from a He WD or a less massive C+O WD. The He 
detonation might then compress enough the core of the WD to induce a second
detonation there, and the mass donor would be flung at its orbital velocity 
and survive. $^{56}$Ni--rich material might be captured by those WDs and its 
decay induce the emission of stellar winds from their surfaces. The WDs, at
times after the explosion comparable with the age of SN 1006, would be hot
UV sources, with luminosities $\sim$ 1 $L_{\odot}$.   

\smallskip
\noindent
Kerzendorf et al. (2018b) have made a deep photometric search of 
the remnant of SN 1006, using the Dark Energy Camera (DECam) of the Dark 
Energy Survey on the 4m Blanco telescope located at Cerro Tololo 
Inter--American Observatory. They compare the observations with both the
predictions from the models of Shen \& Schwab (2017) and with WD cooling
sequences. The latter comparison excludes WDs with cooling ages $\lapprox$ 
10$^{8}$ yr. The observations equally rule out the hot WD models resulting
from radioactive decays taking place at their surfaces (Figure 12).
Kerzendorf et al. (2018) can also rule out most spin--up/spin--down models
as the only possible WD that could escape detection must be older than 
$\approx$ 10$^{8}$ yr. 

\begin{figure}
 \centering
 \includegraphics[width=0.8\columnwidth, angle=0]{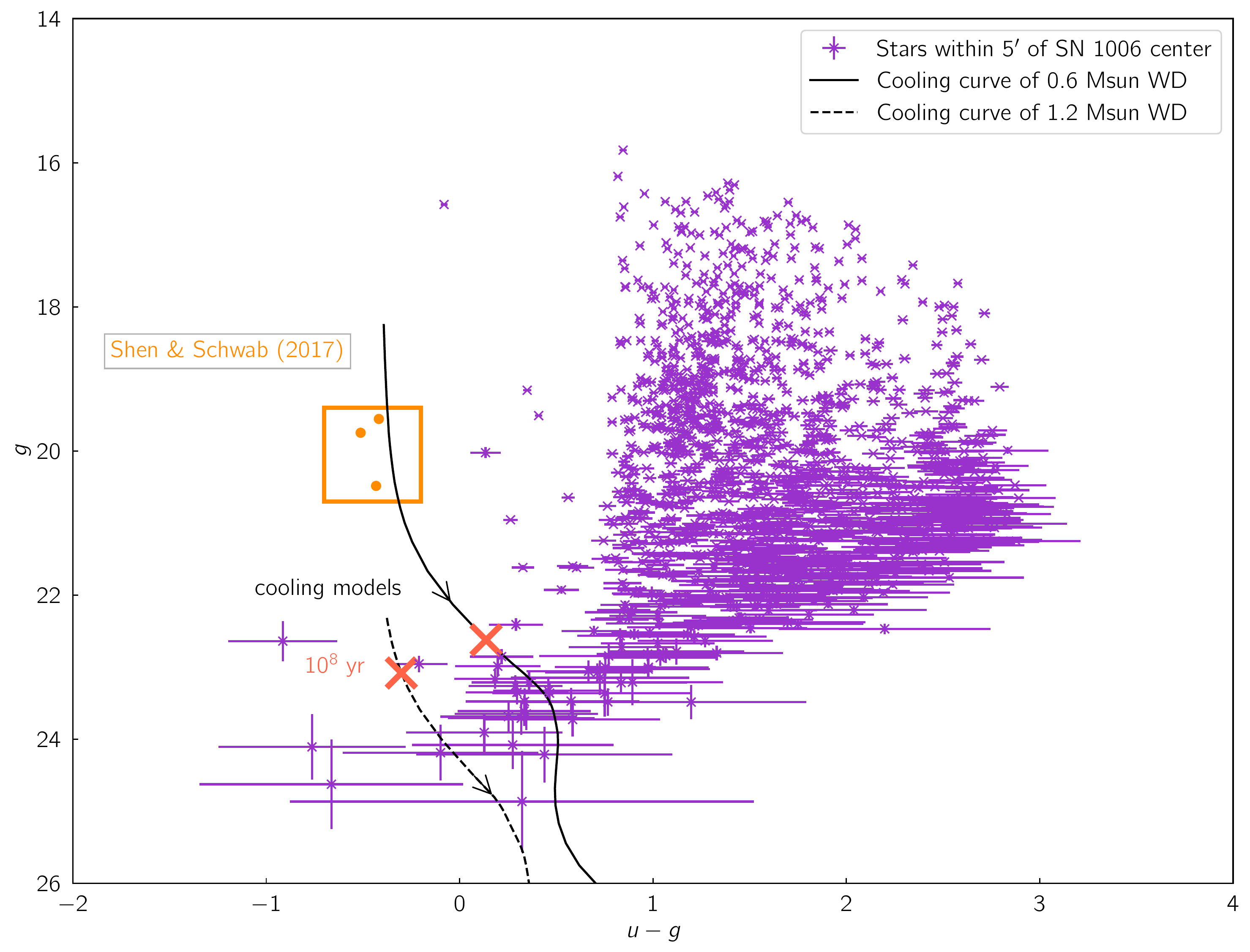}
 \caption{This Figure is taken from Kerzendorf et al. 2017 (Figure 2). WD 
 cooling curves (solid: 0.6 $M_{\odot}$, dashed: 1.2$M_{\odot}$; 
Tremblay et al. 2011). The data are a color--magnitude diagram (CMD) of all 
DECAM stars within 5 arcmin from the center. In orange, the models by Shen 
\& Schwab (2017).
(@MNRAS Permission granted).} 
\label{Figure 9} 
\end{figure}

\subsection{SN 1604 (Kepler's supernova)}

The classification of SN 1604 as SN Ia has been long debated, some authors
adscribing it to the core--collapse  mechanism (Bandiera 1987). X--ray 
observations of the remnant (Cassam--Chena\"{\i} et al. 2004) showed that
the O/Fe ratio was characteristic of SNe Ia (see also Reynoso et al. 2007).  
More recently, Ruiz--Lapuente (2017) has reconstructed, from the historical 
records left by European, Korean and Chinese astronomers, the light curve of
the SN, finding that it was a normal SN Ia.

\smallskip
\noindent
The distance to the remnant of SN 1604 has also been the subject of discussion,
but there is now general agreement that it is $\sim$ 5 kpc (Sankrit et al. 
2016; Ruiz-Lapuente 2017). The remnant appears to be expanding from a center
located at $\alpha_{\rm J2000}$ = 17$^{h}$30$^{m}$41$^{s}$.321$\pm$41$^{s}$.4 and
$\delta_{\rm J2000}$ = -21$^{o}$39'30''.51$\pm$4''.3 (Sato \& Hughes 2017). 
The Galactic latitude being $b$ = 6$^{o}$.8, it lies $\simeq$590 pc above the
Galactic plane. The field is heavily obscured, the extinction being $A_{V}$ 
= 2.7$\pm$0.1 mag (Blair et al. 1991; Schlafly \& Finkbeiner 2011). 

\smallskip
\noindent
Concerning the circumstellar medium of SN 1604, Vink (2008) finds that one
of the components of the binary system that gave rise to the SN might have
created a shell of $\sim$ 1 M$_{\odot}$, expanding into the interstellar 
medium. According to Katsuda et al. (2015) the shell would have lost contact 
with the binary years before the explosion. Chiotellis et al. (2012) and 
Vink (2017) suggest that the companion was an AGB star that had lost its 
envelope at the time of the explosion. 

\smallskip
\noindent
Kerzendorf et al. (2014) made the first exploration of the central region of
Kepler's SNR. The search was photometric and spectroscopic, covering a square
field of 38'' $\times$ 38'' around the center of the remnant determined by 
Katsuda et al. (2008) (only slightly differing, in declination, from the above 
estimate by Sato \& Hughes, 2017). The limiting 
apparent magnitude of the survey was $m_{V} \simeq$ 
18 mag. Given the distance and the extinction, that means reaching a limiting 
luminosity $L \simeq 6\ L_{\odot}$. The WiFeS--spectrograph on the 2.3m 
telescope of the Australian National University was used for the spectroscopy 
and archival {\it HST} images for the photometry.  

\smallskip
\noindent
24 stars were found within the explored area and the magnitude limit of the
survey. From the {\it HST} images, their magnitudes in the F550M filter range 
from 17.24 to 19.24 mag. For 13 of them, there are also V magnitudes taken 
from the {\it Naval Observatory Merged Astrometric Data} ({\it NOMAD}) 
catalogue, that range between 15.9 and 17.7 mag. From Table 1 in Kerzendorf
et al. (2014), the star with the lowest luminosity (in the V filter) in the 
sample had $L = 7\ L_{\odot}$. 
   
\smallskip
\noindent
The quality of the spectra did not allow to infer the stellar atmosphere 
parameters and, as a consequence, no comparison of the absolute magnitudes
corresponding to them with the photometry was possible, so the distances
remained unknown. 

\smallskip
\noindent
Only radial velocities could be measured, for most of the stars (18 of them, 
according with their Table 2), with a typical error of $\approx$ 4.5 km 
s$^{-1}$. No rotational velocity could be determined to better than 200 km 
s$^{-1}$, instead. Kerzendorf et al. (2014) then compared the radial 
velocities found with those given by the Besan\c con model of the Galaxy 
(Robin et al. 2003) for the distribution of such velocities at the distance 
and position of the SNR, and also with the predictions of Han (2008) for the 
velocities of the ejected companions of SNe Ia. They found no star 
significantly deviating from the Besan\c con model. On the other hand, half of 
the stars in the velocity distribution deduced from Han (2008) would appear as 
significant outliers in that model. They did not find any clear candidate
to be the surviving star of the explosion. 

\smallskip
\noindent
From comparison of the observed brightnesses with predicted luminosities, 
Kerzendorf et al. (2014) could discard red giants as possible surviving SN 
companions in Kepler's SNR. That was already an important point, given the 
previous suggestions (see above), from the characteristics of the 
circumstellar medium, that the companion was an AGB star.

\smallskip
\noindent
More recently, Ruiz-Lapuente et al. (2018) have made a new exploration of
the central region of Kepler's SNR. They have surveyed a circle of 24 arcsec
radius around the center of the remnant given by Vink (2008), which is 
practically coincident with that from Sato \& Hughes (2017), down to 
an apparent magnitude $m_{R}  =$ 19 mag which, given the distance and the 
extinction, translates into and absolute magnitude $M_{R} =$ 3.4 mag or a 
luminosity $L = 2.6\ L_{\odot}$. The survey includes spectroscopy with the
multiobject spectrograph {\it FLAMES} on the 8.2m ESO VLT--UT2 and proper 
motions from images taken by the {\it HST}, with a baseline of 10 yr. The 
initial search radius was expanded to 38 arcsec to take advantege of free 
fibers in {\it FLAMES} although the extra stars are too distant from the 
center of the SNR. A total of 32 stars were observed.

\smallskip
\noindent
Proper motions for all 32 stars were measured (Figure 2 and Table 3 in 
Ruiz--Lapuente et al. 2018b). The number of surveyed stars is much larger, 
however, and the astrometric sample can be considered complete down to 
$m_{\rm F814W} \simeq$ 22.5 mag (wide $I$) and 50\% complete down to 
$m_{\rm F814W} \sim$ 23.4 (see the rightmost panel in Figure 3 of that 
paper). 

\smallskip
\noindent
The stellar parameters $T_{\rm eff}$, log $g$, and [Fe/H] were determined from
the spectra obtained. The {\it FLAMES} observations had been made in the 
Combined IFU/7--Fiber simultaneous calibration UVES mode and Giraffe using the 
HR9 and HR15n settings. Two stars were observed both in UVES and Giraffe, thus 
providing a reliability test of the observations. The stellar parameters were
derived from a set of narrow--band spectral indices, following the method
described in Damiani et al. (2014). Then, the distances to the stars were
determined from comparison of the deduced absolute magnitudes $M_{V}$, 
$M_{R}$, $M_{J}$, $M_{H}$, and $M_{K}$ with the photometry of the {\it NOMAD}
catalog, taking into account the corresponding extinctions (Table 5 in 
Ruiz--Lapuente et al. 2018b). The results suggest that the sample is made of
an ordinary mixture of field stars (mostly giants). A few stars seem to have
low metallicities ([Fe/H] $<$ -1) but with large error bars, and they are 
all consistent with being metal--poor giants. 

\smallskip
\noindent
Radial and projected rotational velocities ($v$ sin $i$) were measured from 
the spectra, with errors of 1--2 km s$^{-1}$ in $v_{r}$ and of 10--15 km 
s$^{-1}$ in $v$ sin $i$. The radial velocities are compared with those 
obtained by Kerzendorf et al. (2014), for the stars in common between the 
two samples. There is reasonable agreement in most cases and some unexplained
significant discrepancies in a few of them. 

\smallskip
\noindent
The measured radial velocities and proper motions are plotted over the
corresponding distributions given by the Besan\c con model of the Galaxy,
for the 12 stars at distances shorter than 10 kpc, in Figures 5--7 of 
Ruiz--Lapuente et al. (2018). There are no significant outliers. One star
(T18 in that paper, which is A1 in Kerzendorf et al. 2014) has a very large
proper motion, but it is an M star at a distance of 0.4 kpc only. This distance 
is obtained from stellar modeling and confirmed by the {\it Gaia} DR2 
parallaxes.

\smallskip
\noindent
Ruiz--Lapuente et al. (2018) conclude that from the absence of any peculiar
star down to $\approx$ 2 $L_{\odot}$ and within an angular distance from the
center of the SNR  amounting to 20\% of the average radius of the
remnant, the single--degenerate channel appears clearly disfavoured in the
case of Kepler's SN. There is agreement in that with Kerzendorf et al. 
(2014). Given the characteristics of the circumstellar medium, with a massive
shell expanding from the site of the explosion, the core--degenerate channel
(Kashi \& Soker 2011) appears favored: a WD merging with the also 
electron--degenerate core of a red giant star inside an AGB envelope.

\subsection{RCW86}

RCW86 (also known as G315.4-2.3 or MSH14-63) is the result of a SN explosion 
that is thought to correspond to the ``guest star'' of 185 A.D. observed by 
Chinese astronomers (Clark and Stephenson 1977). This SNR has a radially 
oriented magnetic field similar to those of Tycho, Cas A, and Kepler, which 
confirms its relative youth (Petruk 1999). Its distance is very well known, 
with values comprised between 2.3 kpc (Sollerman et al. 2003) and 2.8 kpc 
(Rosado et al. 1996).

\smallskip
\noindent 
From both radio and X--ray observations, its shape is close to spherical, 
with an angular diameter varying between 40 and 43 arcmin. 

\begin{figure}
\includegraphics[width=0.6\columnwidth,angle=0]{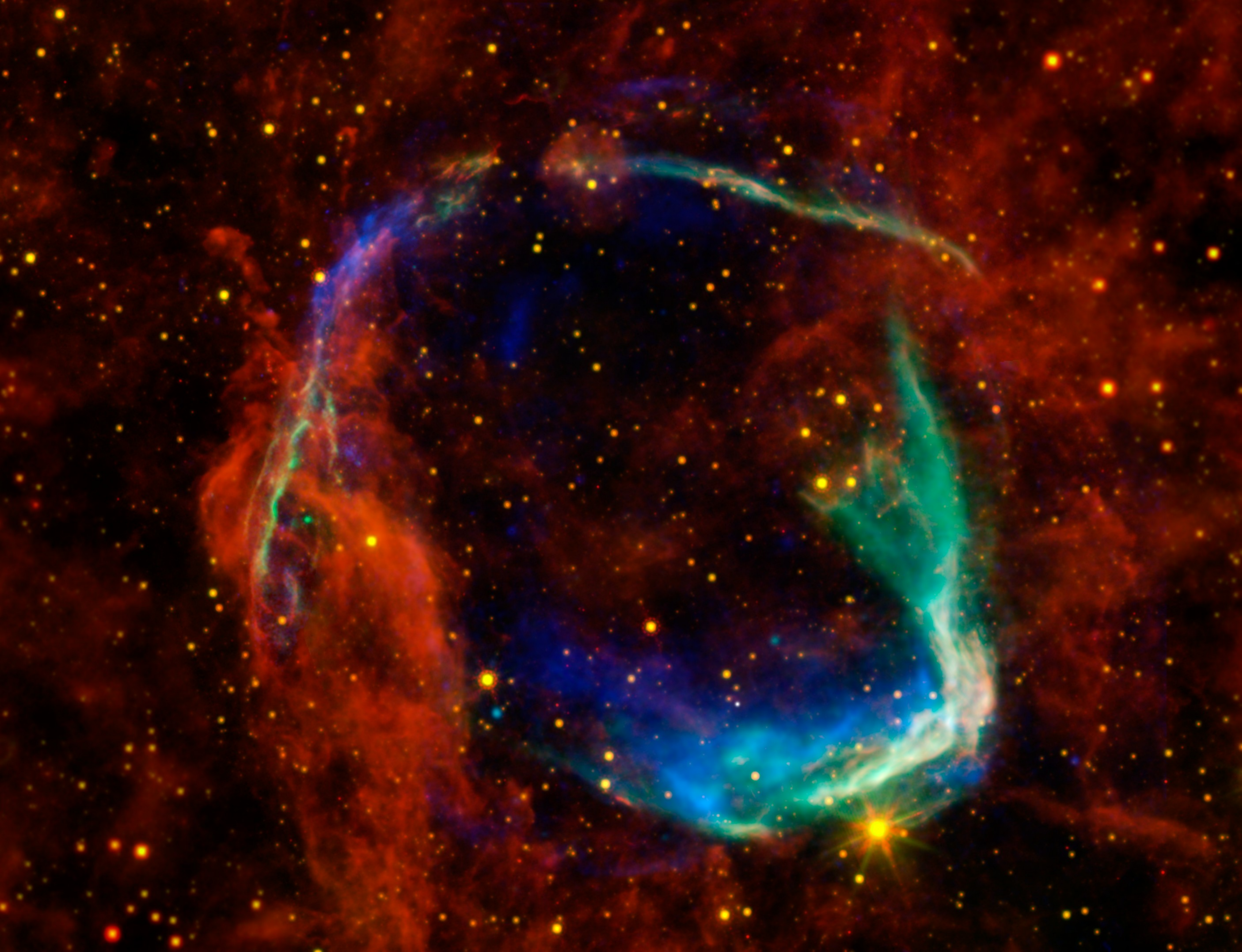}
\caption{Combined image of RCW86, from X--rays ({\it XMM Newton} and {\it 
Chandra}) and infrared data ({\it Spitzer Space Telescope Observatory} and
{\it Wide--Field Infrared Survey Explorer}). X--rays are in blue and green.
Infrared emission, in yellow and red. Public Domain.}
\label{Figure 10}
\end{figure} 

\smallskip
\noindent
An inconvenience is its large diameter:  29 $\pm$ 6 pc, which implies,
at a distance of 2.5 kpc, a 4$'$ radius to enclose 20 $\%$ of the inner
core. More adequate than a 20 $\%$ of the inner core is a 40 $\%$, given
the ill--definition of the remnant (Williams et al. 2011; Lopez et al. 2011; 
see Figure 13).

\subsection{Other Galactic SNe Ia under scrutiny}

\smallskip
\noindent
We now address the steps to clarify the origin of other Galactic SN Ia remnants.
The ejecta of some of these SNe are appreciably asymmetric,
 the diameter of the remnant
being different in the E--W direction than in the the N--S direction.
The origin of the asymmetry is unclear: it might 
either be caused by the explosion mechanism such as 
 off--center ignitions or to double detonation in 
the exploding white dwarf (Maeda et al. 2010; Fink et al. 2010), to expansion 
through a nonuniform medium along the line of sight, or to expansion altered 
by a circumstellar medium modified by planetary nebula--like bipolar outflows
from the companion star of the SN (Tsebrenko \& Soker 2013).

{\it 3C 397} is one of the brightest Galactic SNR in radio and it has an 
irregular shape. Yamaguchi et al. (2015)  presented {\it Suzaku} X--ray
spectroscopic observations detecting high abundances of Ni and Mn. 
They infer in their analysis that this was the explosion of a SN Ia close
to the Chandrasekhar mass. This is found as well in the study by Dave et al.
(2017), that  suggests that this SN Ia falls well into the single degenerate 
scenario, where a central deflagration is induced by accretion of 
mass from a  non--degenerate companion. The deflagration would turn into 
detonation in the WD, according to this analysis.
It is possible to test if there was a companion or not, though this 
requires at the moment a photometric and kinematic 
characterization. The distance of 3C 397 is in the range of 6.3--9.7  kpc and 
its  age is estimated to be around 1350 yr--1750 yr (Leahy \& Ranasinghe 2016). 
It occupies in the sky around $\sim$ 1.7 arcmin in radius.  It is not
 an easy
 search due to the fact that this remnant is on the plane of the Galaxy and
the extinction is very large.  So, it should be postponed to a 
possible examination with the NIRcam on board of the {\it JWST} and give it a  
lower 
priority in front of other remnants.

\begin{figure}
\centering
\includegraphics[width=0.8\columnwidth, angle=0]{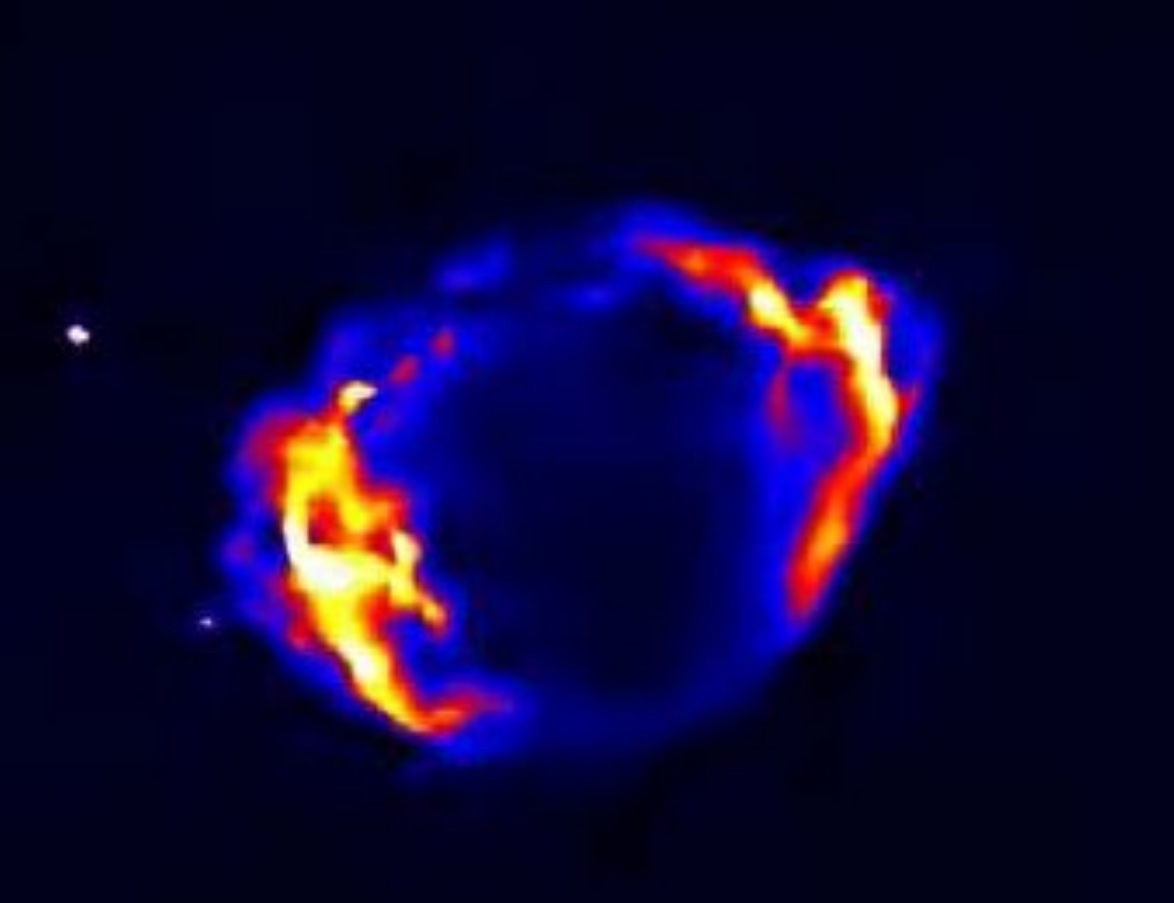}
\caption{Image in X--rays of G1.9+0.3. It is worth reminding that this 
image is of a remnant of only 50 arcsec of radius. The search radius could 
encompass stars within the 10 arcsec radius (20\% of that of the SNR). Image 
adapted from Borkowski et al. (2017). (Courtesy of Kazik Borkowski.
@AAS. Reproduced with permission).}
\label{Figure 11} 
\end{figure}

{\it G344.7-0.1} 
SN remnant was discovered in radio observations (Caswell et al. 1975) and 
has a largely asymmetric structure with a diameter of $\sim$ 10 arcmin.   
X--ray observations made from the {\it Suzaku} satellite (Yamaguchi et al. 
2012) have shown strong K--shell emission from lowly ionized Fe, and the
pattern of abundances is consistent with a SN Ia origin. 
The distance to this SNR is uncertain, a value between 6 and 14 kpc 
(Giacani et al. 2011, Yamaguchi et al. 2012). It seems clear that it 
is located in the opposite edge of the Galacic plane, beyond the Galactic 
center (Yamaguchi et al. 2012). 
 Its age is  estimated to be
between 3000 and 6000 yr. A more detailed X--ray analysis is needed. 
More detailed constrains on its age are required to limit the companion
search in this SN Ia SNR. However, this search is perhaps not proritary 
when compared with other studies. The distance is larger (6--14 kpc) 
than to other remnants and that entails a greater difficulty for gathering 
the information required for the identification of the surviving companion. In 
addition, it is highly extinguished, being very close to the Galactic plane.

{\it G352.7-0.1}
SN remnant is asymmetric, with radius around 3.9 arcmin 
(Sezer \& G\"ok 2014) and elemental abundances that are typical of a SN Ia 
remnant. 
 Its age is $\sim$2200 yr and the distance is estimated to be 
around 7.5 $\pm$ 0.5 kpc by studying the interstellar 
gas surrounding it (Giacani et al. 2009). Its location is very close
to the Galactic plane. The age of the remnant and the shorter distance makes it
a better case  than the previous remnant for 
a survey in the infrared with the {\it JWST}.

{\it G337.2-0.7}. Here again we have an SN Ia remnant
 asymmetric with size of 4.5 $\times$ 5.5 arcmin (diameter).
 The models by Badenes 
et al. (2003) fitting the X--ray spectrum of this remnant give an 
estimate of its age of 5000 years  and an uncertain distance 
between 2--9.3 kpc (Rakowski et al. 2006). This remnant, though
close to the Galactic plane, is a bit higher and has a smaller extinction than 
the three remnants previously addressed. 
Though, from the series of remnants that we are discussing,
 the really prioritary one
for a search for a companion with the {\it JWST} 
is the next remnant that we are about to introduce.

{\it G1.9+0.3} was discovered in 2008, and it is believed to be the remnant 
of a SN Ia which exploded around 1900. The supernova is 8.5 kpc  away 
from us (Carlton et al. 2011) and has only 50 arcsec of radius.  However, it 
is highly extinguished in the optical. Studies of how much would the 
absorption in the infrared be (Reynolds et al. 2008) set an infrared absorption 
of the order of 1.8 mag in the K band.  This is consistent with what is 
obtained from the Besan\c con model of the Galaxy (Robin et al. 2003). The fact 
that this SNR is so recent means that the companion star, even moving at 400 
km s$^{-1}$ perpendicularly to the line of sight, would only be 1.2 arcsec 
away from the site of explosion after the 120 years elapsed since the 
event. Thus a search could include the study of stars within 10 arcsec of 
radius, to allow for the possible difference  between the expanding center of 
the SNR and the explosion site. It could even be enlarged to 30$\%$  and  
that would be 15 arcsec of radius around  the expansion center.
Never before such a small circle of search had been enough for a conservative 
exploration of companion candidates in a SNR in our Galaxy. Thus, this 
SNR offers an exceptional oportunity. One will not have hundreds of 
candidates to examine, as in other cases, but only a reasonable amount 
(see Figure 14).

\smallskip
\noindent
Chakraborti et al. (2016) have studied this very young SNR. They find
that circumstellar interaction in young Galactic SNRs can be used to 
distinguish between SD and DD progenitor scenarios. The
X--ray flux and the size of the remnant are both increasing, and  
these authors introduce a {\it surface brightness index}, relating flux and size
evolution. The theoretical evolution of this index is different in the 
two scenarios, and it is always negative in the SD case (no simultaneous 
increase in flux and size). Since the contrary is observed in the SNR G1.9+03 
case, they favour the DD scenario there: no candidates should be found.

\begin{figure}
 \centering
 \includegraphics[width=0.8\columnwidth, angle=0]{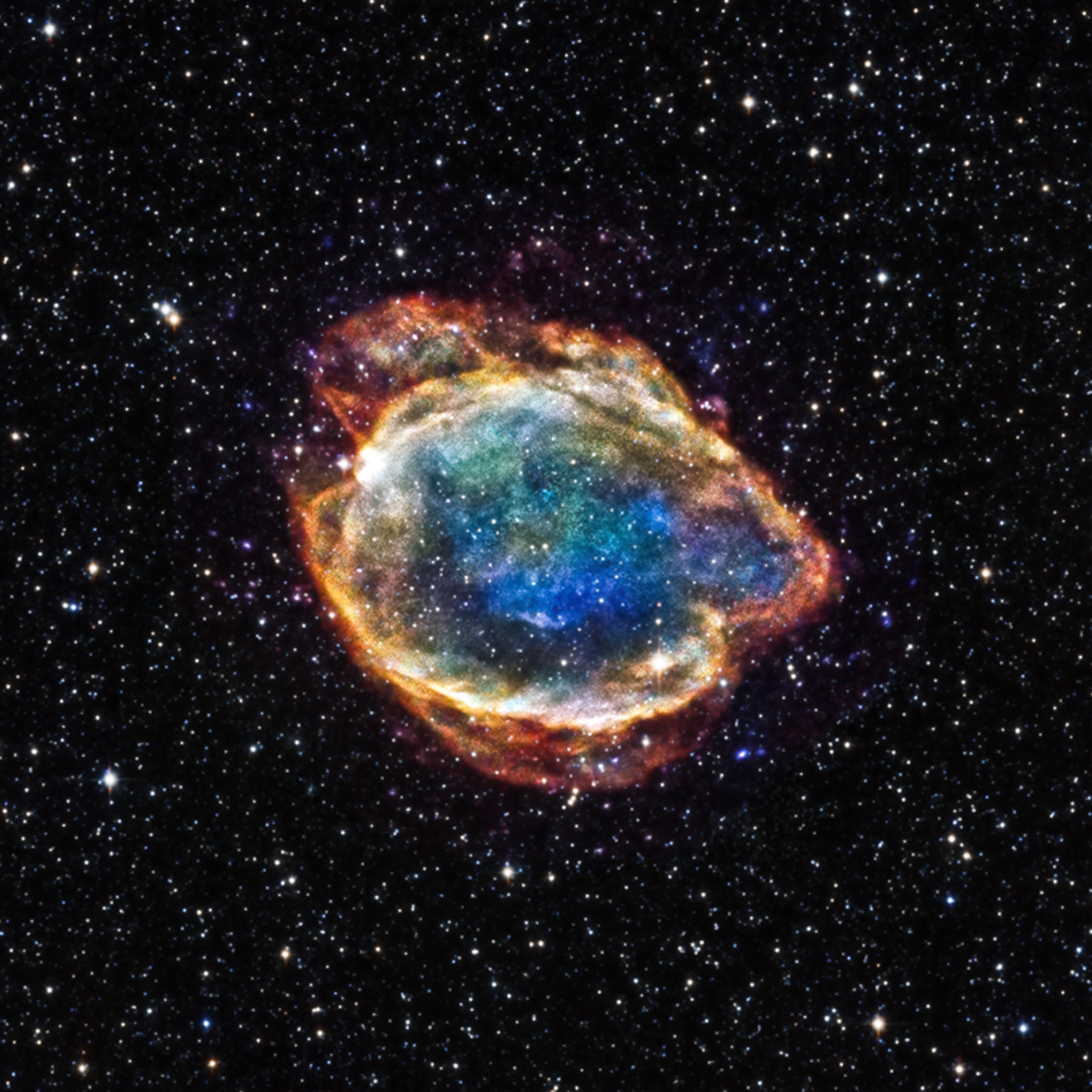}
 \caption{Three--color image of G299.2--2.9, based on {\it Chandra} data. 
Red, green and blue represent
 the 0.4--0.72; 0.72--1.4, and 1.4--3.0 keV bands, respectively.  
 (Courtesy of Sangwook Park.  @AAS. Reproduced with permission).}
 \label{} 
\end{figure}

\smallskip
\noindent
Now, we come to 2 SN Ia remnants that are not highly extinguished and 
nearby. They present a very good opportunity to look for the surviving 
companion of those SNe Ia.

{\it G299.2-2.9}. 
This SNR was discovered, in X--rays, by Busser, Egger \& Aschenbach (1995), 
as a part of the {\it ROSAT All Sky Survey}. 
{\it Chandra} observations have provided measurements of abundance ratios in 
good agreement with the predictions of delayed--detonation models of SNe Ia 
(Post et al. 2014; Park \& Post 2016). The distance 
seems to be $\sim$5 kpc, and the age $\sim$4,500 yr 
(Park et al. 2007; Park \& Post 2016). 
The coordinates of the centroid of the X--ray emission are 
RA = 12$^{h}$ 14$^{m}$ 50$^{s}$.508; Dec = -65$^{o}$ 28' 14''.51 
(J2000) (Post 2017; see Figure 15). 
This corresponds to Galactic coordinates l $=$ 299.142$^{o}$
and b $=$ --2.869$^{o}$. Thus, this SNR is below the Galactic plane. Its
 estimated extinction, $A_{V}$ = 2.3, is moderate (if we take into account the 
whole sample of Galactic SN Ia SNRs) and enables to construct 
color--magnitude diagrams in search of a possible surving companion.  Other 
SN Ia SNRs have been studied with color--magnitude
diagrams, as seen 
in the section on SN 1006 and will be shown in the section on LMC SNRs. Those 
comparisons 
have been centered on companions of the MS and the He star types
 studied by Pan et al. (2014). 
In addition to the classical 
MS, subgiant and red giant companions impacted by the explosion, 
 we would like to extend the searches   
to types of possible surviving companions that have not yet been investigated.  

\smallskip
\noindent
sdB companions would typically have $L \simeq 10\ L_{\odot}$ and $T_{\rm eff}
\simeq 40000 K$. With an intertellar extinction $A_{U}$ = 3.6 mag (Schlafly \& 
Finkelbeiner 2001) and for a distance $\sim$ 5 kpc, we have $m_{U} \simeq$ 
22.5 mag, so going to a $U$ magnitude of 23 should cover these possible hot 
companions, not searched for in any other SNR yet. 
To cover the faint companions up to  $L \simeq 0.01\ L_{\odot}$, 
 with an extinction $A_{V} =$ 2.3 mag, 
we should go, at the distance of the SNR, to $m_{V}$ = 25.7 mag. Therefore,
reaching down to 26 mag in the $V$ and $R$ bands  would be required.

\smallskip
\noindent
{\it G272.2-3.2}. It was discovered 
in the ROSAT All--Sky Survey (Greiner \& Egger 1993), and more recently 
studied by Harrus et al. (2001) and McEntaffer et al. (2013). It
 was produced by 
a SN Ia explosion (Lopez et al. 2011) and it is 6000--12000 yr old.
The centroid position is $\alpha_{\rm J200}$ = 09$^{h}$ 
06$^{m}$ 45$^{s}$.7, $\delta_{\rm J2000}$ = -52$^{o}$ 07' 03'' (Greiner \& Egger 
1993) and l$=$ 272.13$^{o}$,  b$=$--3.19$^{o}$. It is far enough from the plane
of the Galaxy so that, as in the previous case, it
 can be studied through color--
magnitude diagrams. The 
distance is $d = 1.8^{+1.4}_{-0.8}$ kpc (Greiner et al. 1994), or $\sim$ 2--2.5 
kpc according to Harrus et al. (2001) and Kamitsukasa et al. (2016).
 Its 
distance from the Galactic plane, at 2 kpc, would be over 110 pc. The diameter 
of the remnant is slightly less than 20 arcmin.  This SNR has a 
extinction somehow larger than the previous one. Then, 
the estimated requirement to cover the 
colour--magnitude diagrams and 
 to study all the abovementioned possible companions, means to go 
 as deep in magnitude as in the previous remnant G299.2-2.9.

\subsection{Type Ia supernovae in the Large Magellanic Cloud}

\begin{figure}
\centering
\includegraphics[width=0.8\columnwidth,angle=0]{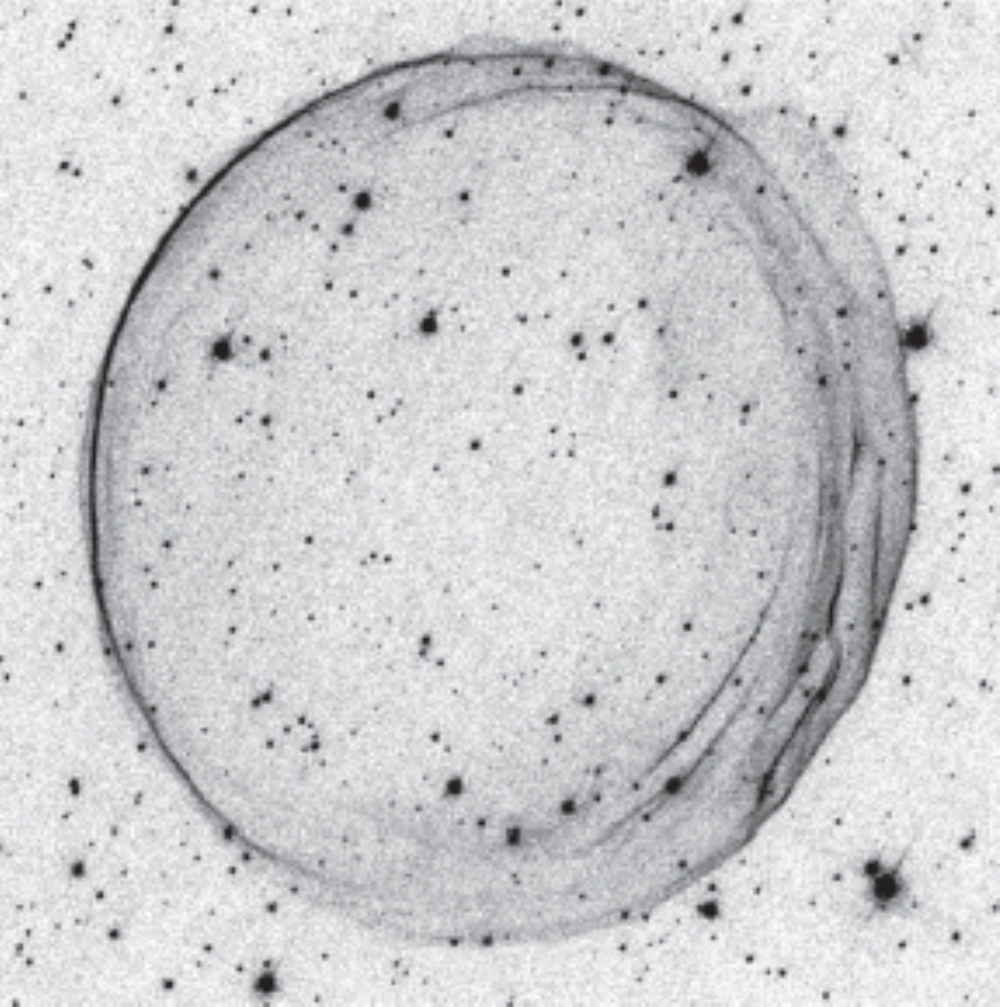}
\caption{SNR 0509-67.5 in the H$\alpha$ filter, obtained by the ACS on board
{\it HST} (from Hovey, Hughes \& Ericksen 2015). (Courtesy of Luke Hovey 
\& Jack Hughes. @AAS. 
Reproduced with permission).} 
\label{Figure 12} 
\end{figure}

Twelve SNRs in the Large Magellanic Cloud (LMC) have been 
identified as being produced by SNe Ia (Litke et al. 2017, and references 
therein). Only five of them have been explored to date.

\smallskip
\noindent
The first remnant to be studied in search for a surviving companion was
SNR 0509-67.5 (Schaefer \& Pagnotta 2012). This SNR was discovered as an 
X--ray source by the {\it Einstein Observatory} (Long et al. 1981) and 
confirmed as a SNR by Tuhoy et al. (1982). Later, an SN Ia origin was 
indicated by {\it Chandra} X--ray observations (Warren \& Hughes 2004) showing
that the ejecta abundances were consistent with the nucleosynthesis 
predictions for delayed--detonation models of SNe Ia. Rest et al. (2005) 
discovered light echoes, based on which the SNR has an age of 400$\pm$120 yr. 
Optical spectra of the light echoes (Rest et al. 2008) showed that the 
explosion was a SN Ia, likely overluminous as SN 1991T.
 
\smallskip
\noindent
From images 
taken with the {\it HST}, no possible companion star was found down to an 
apparent magnitude $V$ = 26.9 mag (corresponding to an absolute magnitude 
$M_{V}$ = +8.4 mag and thus to about 0.04 $L_{\odot}$). The authors explored 
an error circle with 1.43'' radius around the apparent center of the remnant, 
which they deemed sufficient from the estimate of the maximum angular distance 
that the fastest possible companions (main--sequence stars) might have 
travelled in $\approx$ 500 yr (the upper 3$\sigma$ error in the calculated age 
of the SNR). From that, they concluded that the SN should have been produced 
through the DD channel.

\smallskip
\noindent
Such conclusion was challenged by Di Stefano \& Kilic (2012), based on the
spin--up/spin--down mechanism (see above). They argued that the spin--down 
time might have been long enough for the companion to have become a WD, dimmer 
than the limiting magnitude reached in the exploration of Schaefer \& Pagnotta 
(2012). The time scale of spin--down was later addressed by Meng \& 
Podsiadlowski (2013) (see also above), who found that it might have been 
sufficient for a red--giant companion to become faint enough to have escaped 
detection.   

\smallskip
\noindent
Di Stefano \& Kilic (2012) also argued that the region explored might have
been too small, given the possible degree of discrepancy between the 
geometric center of the SNR and the actual site of the explosion.

\smallskip
\noindent
Hovey, Hughes \& Eriksen (2015) (see Figure 16), 
using narrow--band H$\alpha$ images taken
with the {\it HST}, made proper motion measurements of the forward shock 
of the remnant. They found asymmetry in the expansion velocity along an 
approximate E--W axis. Hovey (2016), and Hovey, Hughes \& Eriksen 
(2016), combining these proper motion measurements with hydrodynamical modeling,
calculated the offset of the explosion site from the geometric center of 
the SNR, based on different assumptions. From that, they derived a search 
radius significantly larger than the 1.43'' radius adopted by Schaefer \& 
Pagnotta (2012). Within their new circle they found, from photometry obtained 
with the {\it HST}, no less than 21 stars with I--band magnitudes ranging from 
26.9 to 20.51, which were still to be studied.

\smallskip
\noindent
Litke et al. (2017) have also explored the central region of SNR 0509-67.5, 
adopting an explosion site that differs from that of Schaefer \& Pagnotta 
(2012) 
by 1''.3, and covering a circle of 3'' radius around it (more than twice that
adopted by these authors). By comparing the stars there with the post--impact
explosion models of Pan, Ricker \& Taam (2014), they conclude that no 
normal surviving companion is present. 

\smallskip
\noindent
The second SN Ia to be explored in the LMC was SNR 0519-69.0 (Edwards, 
Pagnotta \& Schaefer 2012). The type of the remnant was assigned from 
the light echo of the SN and also from its X--ray emission. They used 
{\it HST} images with a limiting magnitude $V$ = 26.05 mag. The circle 
explored, based on the same considerations as in the previous case, had a 
radius of 4.7'' from the geometrical center of the SNR. It contained 27 
main--sequence stars brighter than $V$ = 22.7 mag, any of which might have 
been the companion, in the absence of further evidence. There were no 
post--main--sequence stars. The result thus pointed either to a supersoft 
X--ray source as the progenitor system of the SN or to its coming from the DD 
channel. This SNR has later been studied by Li et al. (2019) (se below). 

\smallskip
\noindent
Pagnotta \& Schaefer (2015) observed two further SNRs 
of the SN Ia type in the LMC: SNR 0505-67.9 (also known as DEM L71) and 
SNR 0509-68.7 (or N103B). After locating their centers and tracing the 
corresponding 3$\sigma$ circles, they have found possible candidates of all 
types: 121 stars in SNR 0505-67.9 (among them six red giants and one possible 
subgiant) and 8 stars in SNR 0509-68.7 (N103B). No channel nor type of 
progenitor system could either be confirmed or excluded by these observations.

\smallskip
\noindent
More recently, Li et al. (2017) have also explored the SN Ia remnant 
N103B. The physical structures inside the SNR have been 
studied from H$\alpha$ and continuum images obtained with the {\it HST} and 
with high--dispersion spectra taken at the 4m and 1.5m telescopes of the Cerro 
Tololo Inter--American Observatory. After determining the explosion center, 
they have found, close to it, a star whose colors and luminosity are 
consistent with a 1 $M_{\odot}$ subgiant companion concordant with a model by 
Podsiadlowski (2003). In this model, the star has a 0.2 $M_{\odot}$ of 
envelope stripped and the rest is heated by the impact of the SN ejecta. No 
observations allowing to measure radial velocities, rotation or proper 
motions are yet available to either confirm or reject the proposed 
identification.

\smallskip
\noindent
Li et al.(2019), in a new study of SN Ia remnants in the LMC, have 
examined three of them: 0519-69.0, 0505-67.9 (DEM L71), and 0548-70.4. New 
images of SNR 0519-69.0 and SNR 0548-70.4 have been obtained using 
the UVIS channel of the Wide Field Camera 3 (WFC3). SNR 0519-69.0, DEML71 and 
SNR 0548–70.4 were observed with the Advanced CCD Imaging Spectrometer (ACIS) 
of the {\it Chandra} X--ray Observatory. Multi-Unit Spectroscopic Explorer 
(MUSE) observations of SNR 0519-69.0 and DEM L71 were obtained with the Very 
Large Telescope (VLT) of the European Southern Observatory. Stellar 
parameters and  radial velocities were derived from the spectra. 

\smallskip
\noindent
Li et al. (2019) have found a star in 0519-69.0 and another one in DEM L71 
that move at high radial velocities, more than 2$\sigma$ from the mean radial 
velocity of the populations explored. A photometrically peculiar star has been 
found in SNR 0548-70.4, but it might in fact be a background galaxy.  

\smallskip
\noindent
To search for possible surviving companions of the SNe Ia, Li et al. (2019) 
use two different methods. In the first one, the photometric measurements with 
the {\it HST}, of the stars around the center of the SNR, are used to construct
color--magnitude diagrams (see Figure 17) and compare positions of the stars in the diagrams
with the theoretical expectations of the post--impact evolution of MS and 
He--star companions. In the second method, they use spectroscopy with the 
Multi--Unit Spectroscopic Explorer ({\it MUSE}), at the ESO {\it VLT}, to find 
peculiar radial velocities as a mark of possible companions. 

\smallskip
\noindent
In the photometric approach, the calculations of the post--impact evolution 
in luminosity and effective temperature of surviving SN Ia companions of Pan, 
Ricker \& Taam (2014) are used (as in Litke et al. 2017) to construct the
theoretical color--magnitude diagrams  that are 
compared with the observations. Only stars with $V <$ 22.7 mag were 
considered. Any star falling on an evolutionary track whose age (that of the 
SNR) were consistent would be a companion candidate. 

\smallskip
\noindent
In the spectroscopic approach, the physical parameters and radial velocities 
of the stars within the search radius from the site of the explosion are 
determined. Here the magnitude limit is $V <$ 21.6 mag. The photometric 
candidates brighter than this limit are then checked for large radial 
velocities as diagnostics of surviving companions. 

\smallskip
\noindent
SNR 0519-69.0 had previously been studied by Edwards, 
Pagnotta \& Schaefer (2102), but the photometry of Li et al. (2019) is more
complete, including the $B$, $V$, $I$ and H$\alpha$ bands. Taking into account 
both the uncertainty on the center 
and the maximum runaway velocity of a possible He--star companion (the
fastest moving case), Li et al. (2019) consider all stars with $V <$ 23.0 
within a search radius of 2''.7 (corresponding to 0.65 pc at the distance
of the LMC). That includes only eight stars. Their colour--magnitude diagrams 
are compared with the post--impact evolutionary tracks of
MS and He stars. The authors conclude on the absence of 
viable candidates. Only five of these eight stars have $V <$ 21.6 mag, and 
their
stellar parameters and radial velocities have been determined. 
A radial velocity distribution has been obtained 
by including stars within a much larger distance from the explosion site. Only 
one star, moving at 182$\pm$0 km s$^{-1}$, deviates more than 2.5$\sigma$ from 
the mean velocity of the extended sample (264 km s$^{-1}$).

\begin{figure}
\centering
\includegraphics[width=0.8\columnwidth, angle=0]{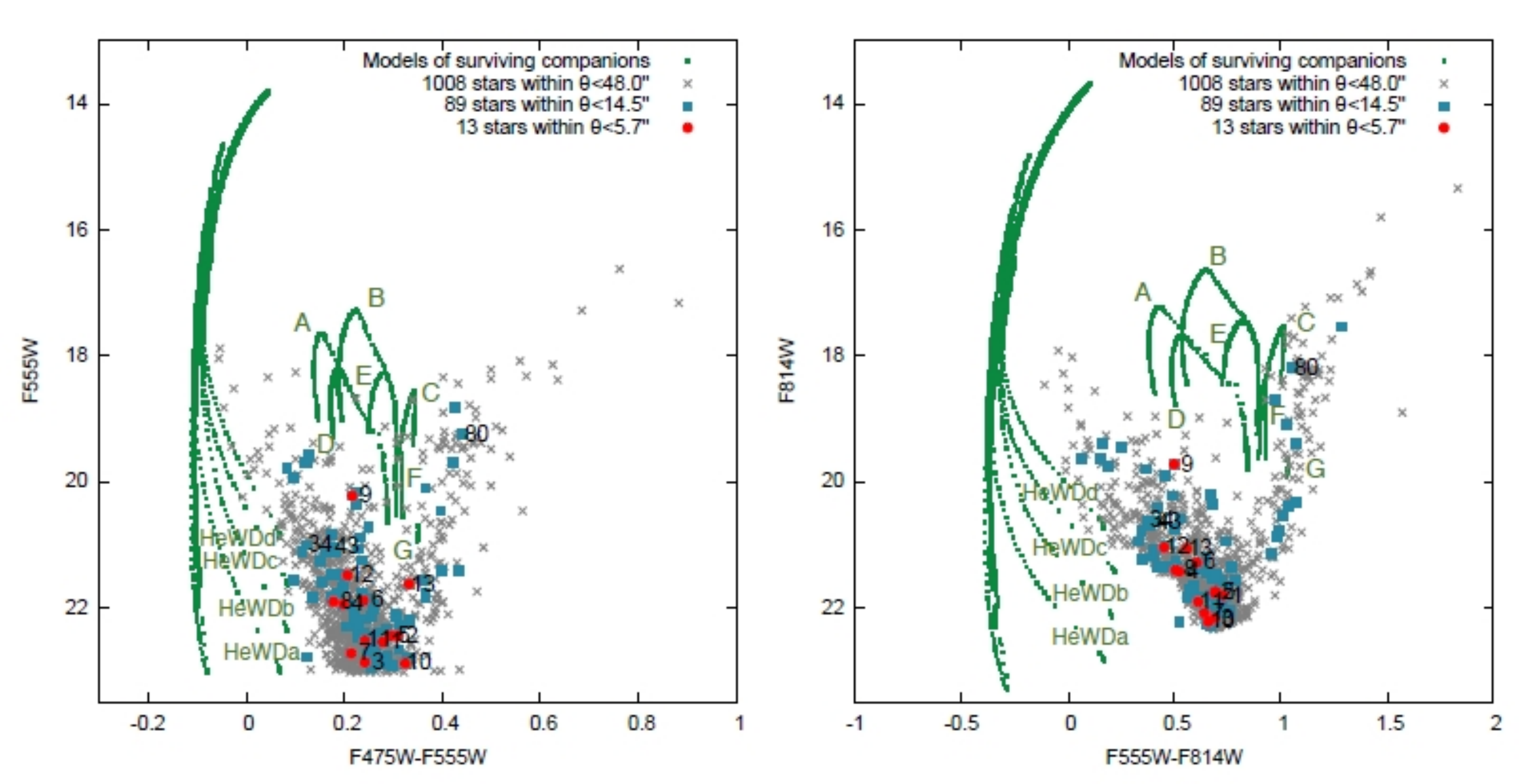}
\caption{Left:  V versus B--V colour--magnitude diagrams (CMD) 
of DEM L71. Right: I versus V--I CMD of DEM L71. Stars that are found 
within the runaway distances from the center for
He--star and main--sequence (MS) surviving
companions, are plotted as blue filled squares and 
red filled squares, respectively.  Stars that are 
superposed on or near the remnant are plotted as grey crosses to 
illustrate  the general background stellar population. The post--impacted 
evolutionary tracks are plotted as small green squares, and those of 
surviving He star and MS companions are to the left and above the MS, 
respectively. Different tracks of He stars and MS companions correspond 
to different companion masses in a range of 0.70 to 1.21 M$_{\odot}$ and 
1.17 to 1.88 M$_{\odot}$, respectively. The details of these He star and
MS companions can be found in models from Pan, Ricker \& Taam (2014). 
(Image and caption as in Li et al. 2019. Courtesy of 
W. Kerzendorf. @AAS. 
Reproduced with permission.)}
\label{Figure 12} 
\end{figure}

\smallskip
\noindent
DEM L71 had been studied by Pagnotta \& Schaefer (2015). The search radius is 
of 14''.5, which corresponds to 3.7 pc. There is a total of 89 stars within that
radius, with $V <$ 23.0 mag. As in the case of SNR 0519-69.0, the theoretical 
evolutionary paths for surviving MS and He--star companions are overplotted
on the observed color--magnitude diagrams. The 
comparison does not yield any candidate. There are 32 stars with $V <$ 21.6 
mag within the 14''.5 radius, for which stellar parameters and radial 
velocities have been determined. When plotting the distribution 
of radial velocities, one star is at more than 2.5$\sigma$ from the mean 
(213$\pm$0 km s$^{-1}$ and 270 km s$^{-1}$, respectively). 

\smallskip
\noindent
SNR 0548-70.4 is explored for the first time by Li et al. (2019). Only the 
photometric approach has been possible. The search radius is 
of 40''.0, corresponding to 10.0 pc. There are 973 stars with $V <$ 23.0 mag
there. Comparison of the color--magnitude diagrams with the 
theoretical evolutionary tracks does not reveal any promising candidate, but 
nevertheless one star has unusual photometric characteristics. Its colors are 
inconsistent with any spectral type. There is the
possibility, however, that the object were not a star but a background galaxy.
Infrared photometry, extending the observed spectral energy distribution,
could discriminate between the two possibilities. 

\smallskip
\noindent
Summarizing, out of the five SN Ia remnants in the LMC that have already been 
explored to some extent, stars with radial velocities that are significantly 
higher than the mean for the corresponding population have been found in two
of them (SNRs 0519-69.0 and DEM L71). A star with characteristics similar 
to those predicted by a model of the post--impact evolution of a surviving
companion appears close to the center of SNR N103B, and a photometrically
peculiar star in SNR 0548-70.4. Nothing has been found in SNR 0509-67.5. 
Further 
observations are needed to measure proper motions, rotational velocities and
chemical abundances of the candidates found in SNRs 0519-69.0, DEM L71 and 
N103B (also the radial velocities, in the latter case), and to elucidate the
true nature of the object in SNR 0548-70.4. Seven more, already 
identified remnants, remain to be explored.

\section{Summary and conclusions}

SNe Ia can, in principle, be produced through two different channels: the SD
channel and the DD channel (the core--degenerate CD scenario, the merging of a 
WD with the electron--degenerate core of a red--giant star, being a variant 
of the DD channel). We still do not know, at present, in which proportions 
(including zero) does each channel contribute to the observed SNe Ia rate. 
Anyway, we can confidently discard that all SNe Ia would come from the 
classical SD channel.  
   
\smallskip
\noindent
In SNe Ia produced through the SD channel, the companion stars of the WDs
that explode should, in general, survive the explosions and thus be detectable
in deep enough surveys. In the case of recent SNe Ia, such companions cannot
have travelled far from the site of the explosion and must, therefore, still
be in the central regions of the corresponding SNRs. Their clear absence there 
would be proof that the SN was produced through some variant of the DD channel.

\smallskip
\noindent
In the last 15 years, several surveys have been made of SNRs of the Ia type, 
in our Galaxy and in the LMC, and others are in progress or have been planned.
The remnants of the four ``historical'' SNe Ia (happened within the last 
2,000 yr and for which we have records of their observations) have been 
explored in some depth, combining ground--based observations with others 
made with the {\it HST}. No clear--cut evidence of a surviving companion has 
been found in any of them, the case of a proposed companion in Tycho's SNR
remaining debated. In the case of SN 1604 (Kepler's SN), the evidence
excludes the SD scenario (Ruiz--Lapuente et al. 2018 suggest the CD scenario),  and in that of SN 1006 a merging of two WDs origin
 is favored.

\smallskip
\noindent
Up to now, five young SNe Ia remnants in the LMC have been explored to some 
extent, by means of the {\it HST} and of {\it Chandra} and other X--ray 
observatories, 
plus radio and other ground--based observations: SNR 0509--67.5, SNR 
0519--69.0, 
SNR 0505--67.9 (or DEM L71), SNR 0509--68.7 (or N103B), and SNR 0548--70.4. 
In the first one, a DD origin (WD+WD merging) seems favored, but the  searches 
might need to go deeper in magnitude to exclude the very faint companions.
 In the other 
four, there is one suggested companion (in SNR 0509--68.7/N103B), 
post--main--sequence companions are excluded in SNR 0519--69.0, and 
 main--sequence 
companions in SNR 0505--67.9, and  
SN 0548--70.4 are being  examined.

\smallskip
\noindent
Thus far, the statistics (although still based on small numbers) seem to 
disfavor the SD channel. One source of uncertainty, however, is to which 
extent the spin--up/spin--down mechanism could be responsible for the failure 
to detect companions due to their faintness. If spin--down times are typically
$t \approx 10^{7}$ yr and the absence of companions would persist in larger 
samples of remnants, the spin--up/spin--down mechanism might hardly be 
invoked as an explanation any longer. In the case of SN 1006, the 
spin--up/spin--down  
may arguably be  disfavoured by the observations of  Kerzendorf 
et al. (2018b), that cover the presence of faint companions and found no
evidence of dim ones, 
although this needs to be checked with hydrodynamical
simulations of the supernova impact on very low mass stars.

\smallskip
\noindent
Another point affecting the searches is the fact that the actual sites of
the explosions may differ from the present centroids of the SNRs by a
considerable angular distance, even in remnants that appear round. That 
should be taken into account in the searches, in addition to the angular 
distance a 
companion might have travelled in the time between the explosion and the 
observations.  

\smallskip
\noindent
Several already identified Galactic SNRs of the Ia type remain to 
be explored
and a larger number in the LMC.
 So, the statistics can rapidly improve.
In the future, not only new ground--based telescopes and the  {\it HST} will be 
of use, but also the {\it JWST} could play an important role, as well as  the  
astrometric data from the upcoming releases by the {\it Gaia} space mission. 
 Some 
surviving companions must eventually be detected, unless the SD channel were
not happening in nature. 

\bigskip
\noindent
\section{Acknowledgements}
\noindent
The author would like to thank the coauthors of her papers related to the 
present review for sharing this exciting endeavour. She  thanks the 
 anonymous referee for his/her comments, that have been incorporated in the 
paper. Special thanks go to
Wolfgang Kerzenzdorf for valuable comments, and suggestions on 
the draft, as well as for allowing the reproduction of various figures 
displayed in papers by him and his collaborators.
 Thanks go as well to the rest of authors of figures reproduced 
here with permission. P.R.--L. is supported by funds 
AYA2015--67854--P  and PGC2018--095157--B--100
 from the Ministry of Industry, Science and Innovation of Spain and the FEDER 
funds.

\end{document}